\documentclass[aps,prd,preprintnumbers,twocolumn]{revtex4-2}

\usepackage{graphicx}
\usepackage{amsmath,amssymb}
\usepackage{float}
\usepackage{xcolor}
\usepackage{multirow}
\usepackage{wrapfig} %% for figure
\usepackage{bm}
\usepackage{siunitx}
\newcommand{\s}[1]{\scriptscriptstyle #1}

\usepackage{hyperref}
\hypersetup{
    colorlinks=true,     % false: boxed links; true: colored links
    linkcolor=blue,      % color of internal links
    citecolor=blue,      % color of links to bibliography
    filecolor=blue,      % color of file links
    urlcolor=blue        % color of external links
}

\usepackage[nameinlink]{cleveref}

  \def\gsim{\mathrel{\rlap{\lower0.25em\hbox{$\sim$}}\raise0.2em\hbox{$>$}}} 
  \def\lsim{\mathrel{\rlap{\lower0.25em\hbox{$\sim$}}\raise0.2em\hbox{$<$}}}
  \def\lg{\mathrel{\rlap{\lower0.25em\hbox{$>$}}\raise0.25em\hbox{$<$}}}
  \def\gl{\mathrel{\rlap{\lower0.25em\hbox{$<$}}\raise0.25em\hbox{$>$}}}

\begin{document}

\title{Thermal Static Potential and Pseudo-Scalar Quarkonium Spectral Functions from 2+1 Flavor Lattice QCD}

\author{Sajid Ali$^{a,b}$, Dibyendu Bala$^a$, Olaf Kaczmarek$^a$, Pavan$^{a}$}

\affiliation{
$^a$Fakult\"at f\"ur Physik, Universit\"at Bielefeld, D-33615 Bielefeld, Germany\\
$^b$Government College University Lahore, Department of Physics, Lahore
54000, Pakistan
}

\collaboration{HotQCD Collaboration}

\date{\today}

\begin{abstract}
Quarkonia, which are bound states of a heavy quark and antiquark, play a key role in probing the quark-gluon plasma (QGP). The dynamics of quarkonia in the QGP are encoded in their finite-temperature spectral functions. In this work, we estimate the quarkonium spectral functions in the pseudo-scalar channel using 2+1 flavor lattice QCD with a pion mass of $320\,\text{MeV}$, at temperatures of $220\,\text{MeV}\,(1.2\,T_{pc}),\,251\,\text{MeV}\,(1.4\,T_{pc})\,\text{and}\,293\,\text{MeV}\,(1.6\,T_{pc})$.
Reconstructing the spectral function from the Euclidean lattice correlator is a well-known ill-posed problem, requiring additional physics-motivated input.
We address this by smoothly matching contributions from different frequency regions of the spectral function, using appropriate physics valid for each region. 
The spectral function around $\omega \sim 2\,M_q$ is obtained using a non-perturbative complex potential, while for $\omega \gg 2\,M_q$ it is modeled using results from vacuum perturbation theory. Since the pseudoscalar channel does not receive a transport contribution near $\omega \sim 0$, we find that the combination of these two regions already provides a good description of the relativistic lattice pseudoscalar correlator.
We observe a substantial thermal width in the $\eta_c(1S)$ state, indicating that pseudoscalar charmonium ($\eta_c$) is nearing dissolution at the studied temperatures. In comparison, the $\eta_b$ ground state exhibits little change and remains well-defined.
\end{abstract}

\maketitle

\section{Introduction}
\label{sec:intro}
Lattice QCD calculations indicate that hadronic matter at high temperatures transforms into the Quark-Gluon Plasma (QGP). This phase is studied in heavy-ion collision experiments at RHIC and LHC. Quarkonia (bound states of a quark and an antiquark) are important probes to study this phase \cite{Matsui:1986dk}. In heavy-ion collisions, quarkonia are generated  in the early phase of the collision through hard processes. As they move through the plasma, some dissociate due to thermal interactions with plasma constituents, while others persist and decay later in the hadronic phase after freeze-out. The heavy quark anti-quark vector mesons ($\Upsilon$, $J/\Psi$) are particularly significant due to their decay into dileptons. Experimental data reveal a suppression of $\Upsilon$ yields in Pb-Pb collisions compared to expectations from proton-proton collisions, indicating significant effect of the QGP on quarkonium states \cite{CMS:2018zza}.

Understanding the effect of the QGP on quarkonia theoretically, requires the real-time evolution of quarkonium states within the QGP, together with the hydrodynamic evolution of the QGP itself \cite{Krouppa:2017jlg,Dong:2022mbo}. Significant progress has been made in describing the real-time evolution in the weak coupling regime using potential non-relativistic QCD (pNRQCD) together with the open quantum system approach. In this framework, one evolves the density matrix of the $Q\bar{Q}$ pair in the singlet and octet representations using Lindblad equations \cite{Kajimoto:2017rel, Brambilla:2022ynh, Brambilla:2023hkw, Sharma:2019xum}. Recently, the quarkonia evolution in the open quantum system approach has been extended to a non-Markovian process to quantify the memory effect \cite{R:2025qoq}.

On the other hand, one can also study the fate of quarkonium bound states and their real-time properties at a given temperature by studying the quarkonium spectral functions across different quantum number channels. The hydrodynamic evolution of the QGP assumes an (approximate) state of local thermodynamic equilibrium. Therefore, understanding the spectral properties of quarkonia at a given temperature provides important insight into heavy-ion phenomenology. The spectral function is defined by,
\begin{equation}
     \rho_{\s{\Gamma}}(\omega,\vec k)=\int dt\, d^{3}\vec x \,\exp[i(\vec k.\vec x-\omega t)]\,\langle [J_{\s{\Gamma}}(\vec  x,t), J_{\s{\Gamma}}(\vec 0,0)]\rangle_{\s{T}} \; .
\end{equation}

Here $\Gamma$ specify the quantum channel of the point current, $J_{\s{\Gamma}}=\bar \psi \Gamma \psi$ and  $\langle ... \rangle_{\s{T}}$ denotes the thermal average. Physically, the spectral function is related to the imaginary part of the retarded response function \( \mathcal{G}_{R}(\omega, \vec{k}) \), which represents the ability of the system to absorb energy due to a perturbation with frequency \( \omega \) and momentum \( \vec{k} \). Spectral function is related to the Euclidean correlation function by:
\begin{equation}
	G^{\s{E}}_{\s{\Gamma}}(\tau, \vec k)=\int_{0}^{\infty} \frac{d\omega}{\pi} \rho_{\s{\Gamma}} (\omega,\vec k)\, \frac{\cosh[\omega(\tau-\frac{1}{2T})]}{\sinh[\frac{\omega}{2T}]} \; ,
\label{illcond}
\end{equation}
where $G^{\s{E}}(\tau ,\vec k)=\int d^3 \vec x \,\exp(i\vec k.\vec x) \langle J_{\s{\Gamma}}(\vec  x,\tau) J_{\s{\Gamma}}(\vec 0,0) \rangle$ and which can be determined in first principle lattice QCD calculations. However, inverting the above equation for spectral function is a well-known ill-posed problem, as lattice QCD data involve a finite number of data points along with statistical errors. This results in a large number of possible spectral functions that agree with the lattice data within errors and further physics input is required for a reliable spectral function reconstruction. There are various studies on quarkonium spectral functions using lattice correlator. Bayesian methods, like the Maximum Entropy Method (MEM), have been used in \cite{Asakawa:2003re,Datta:2003ww,Ding:2012sp} to investigate charmonium in plasma using quenched gauge field background. Similar techniques have been used to study correlators in NRQCD in background gauge configuration with dynamical  HISQ fermion and Wilson-Clover fermion  \cite{Kim:2018yhk,Aarts:2014cda}. While these studies provide insights into the properties of the qurkonia in plasma, they are accompanied by significant systematic uncertainties, which are inherent to Bayesian analysis. There are also other studies of quarkonia from lattice QCD, where various models of the spectral function have been assumed to fit lattice QCD data \cite{Larsen:2019bwy,Larsen:2019zqv,Ding:2025fvo}. For the study of  quarkonia from the T-matrix approach see \cite{Liu:2017qah, Tang:2024dkz, Tang:2025ypa}.

Using non-relativistic approximations for quarkonium bound states various potential models have also been used, where the Schr\"odinger  equation is solved to obtain the spectral function.  In \cite{Mocsy:2007yj}, both the free energy \cite{Kaczmarek:2002mc} and internal energy of the $Q\bar Q$ system were used to study the spectral functions. However, the precise definition of the potential, which governs the real-time dynamics of the heavy $Q\bar Q$ system, has been clarified in \cite{Laine:2006ns}. This was achieved through a rigorous $1/M_q$ expansion of the point-split version of the correlator, resulting in a Schr\"odinger equation with a thermal potential. Perturbatively, this calculation revealed for the first time that the potential at finite temperature is complex. The real part corresponds to the Debye-screened potential and the imaginary part is due to Landau damping in the  $r\gg \frac{1}{T}$ region, which arises from the inelastic scattering of medium quarks and gluons with the heavy quark. Later, within the pNRQCD framework, it was shown that in the region $r \ll \frac{1}{T}$, the imaginary part of the potential receives additional contributions from singlet-to-octet transitions due to the absorption of a gluon from the medium (known as gluodissociation) \cite{Brambilla:2008cx}.

This perturbative potential has been used for the spectral function calculation in various channels near the threshold \cite{Laine:2007gj, Burnier:2007qm}. This illustrates that the imaginary part plays a significant role in the dissociation of quarkonia in the QGP, along with the color screening of the real part. 
These perturbative calculations are important inputs for spectral reconstruction of the lattice correlator \cite{Burnier:2017bod, Ding:2021ise}. Some of the results presented in this paper have already appeared in preliminary form in the proceedings \cite{Bala:2024keu}. The spectral function in the entire frequency range was constructed by combining the result from the threshold region, based on the thermal potential, with spectral functions from other regions. Using these model spectral functions, the corresponding perturbative correlators have been computed using \Cref{illcond} and have then been compared to the lattice correlators. The lattice data was fitted by introducing some free parameters in the perturbative model spectral function. 

In this paper, we used a similar approach for the spectral reconstruction, with one key difference: in the threshold region, we use a non-perturbative thermal complex potential to obtain the quarkonium spectral functions. This is particularly relevant for temperatures just a few times higher than the pseudo-critical temperature $T_{pc}$, where non-perturbative effects in the potential may play a significant role. There are also existing spectral function calculations based on non-perturbative potential \cite{Burnier:2015tda}. In this work we combine spectral functions from different regions with a more precise extraction of the thermal potential using data from large \( N_\tau \) lattices. We focus on the pseudo-scalar channel spectral function as in this channel no transport contribution is present in the infrared region. As a result, the large-distance part of the correlator is dominated by the spectral contribution from the bound state region. This enables us to compare the correlator obtained from the spectral function with the one directly calculated from point-to-point lattice QCD correlators.

However, calculating the non-perturbative potential itself turns out to be an inverse problem, where the correlator is related to the Wilson loop. Over the decade, significant efforts have been devoted to compute the thermal static potential non-perturbatively. These approaches range from Bayesian techniques like the Maximum Entropy Method (MEM) \cite{Rothkopf:2011db} and Bayesian Reconstruction \cite{Burnier:2014ssa} to models of spectral function that are fitted to the lattice data \cite{Bala:2019cqu, Bala:2021fkm,Bazavov:2023dci,Larsen:2024wgw}. Some models even suggest that there is no screening of the real part of the potential \cite{ Bala:2021fkm, Bazavov:2023dci}. However, given the ill-conditioned nature of this inverse problem, this could very well be due to the lack of physics information used during the spectral reconstruction. In fact, as we will see, and as also discussed in \cite{Bala:2019cqu}, incorporating physics-motivated information into the spectral function results in a color-screened thermal potential.

The organization of the paper is as follows: In the next section, we discuss the formalism for calculating the spectral function using the thermal static potential. In \Cref{Lattice}, we provide details of our lattice simulations. \Cref{potential} covers the detailed analysis of the thermal potential extraction. In \Cref{Pole_mas}, we determine the quark pole mass and the overall additive constant of the thermal potential, and we also provide a parametrization of the thermal potentials. In \Cref{Spec fn}, we use the thermal potential to extract the spectral function in the threshold region and also match it with the UV perturbative spectral function to obtain the spectral function over the entire region. In \Cref{check}, we perform a consistency check to show that the spectral function is compatible with the corresponding lattice correlator at finite temperatures, and we also discuss the physical implications of our results. The paper end with a conclusion and outlook.

\section{Formalism}
\label{Formalism}
We aim to reconstruct the spectral function using lattice correlators where the \(Q\bar{Q}\) state is in the pseudoscalar (\(\eta_c, \eta_b\)) channel. This leads us to consider the following lattice correlators:

\begin{align}
G^{\s{E}}_{\s{PS}}(\tau) &= M_{\s{B}}^2 \int d^{3}\vec{x} \, \langle \bar{\psi}(\vec{x}, \tau) \gamma_{5} \psi(\vec{x}, \tau) \, \bar{\psi}(\vec{0}, 0) \gamma_{5} \psi(\vec{0}, 0) \rangle_c, 
\label{def.pt}
\end{align}

where \(\psi\) represents the heavy quark field on the lattice. The pseudoscalar correlator is multiplied by the square of the bare quark mass, \(M_{\s{B}}^2\), to ensure that the correlator is ultraviolet (UV) finite after renormalization \cite{Burnier:2017bod}.

It is important to note that on the lattice, this correlator has finite renormalization factors, and a lattice spacing-dependent renormalization factor must be multiplied to make the correlators finite in the continuum limit. However, in this paper, we foucs on a single lattice spacing and estimate the spectral function up to some multiplicative constant. Therefore, we do not need to consider such renormalization for our purposes.

We reconstruct the spectral function using physics-motivated input according to \cite{Burnier:2017bod}. 
In the region where \(\omega \gg 2 M_q\), perturbative calculations become reliable. In this regime, temperature effects are also suppressed because the kinetic energy of the produced quark-antiquark pair is much larger than the thermal energy. Perturbative calculations show that at leading order (LO), temperature effects are exponentially suppressed, and at next-to-leading order (NLO), they decrease according to a power law \cite{Caron-Huot:2009ypo}. Thus, in this region, we estimate the spectral function using vacuum perturbation theory.

Around the threshold region $\omega \sim 2\,M_q$, naive perturbative calculations break down. In this region, The spectral function follows from the solution of a Schr\"odinger equation. This is equivalent to obtaining the spectral function in NRQCD at the order $1/M_q$ expansion. The thermal potential arises from resumming the infinite set of ladder diagrams near $\omega \sim 2\,M_q$ \cite{Burnier:2007qm}. This requires solving the Schr\"odinger equation:

\begin{equation}
\left(2M_{\s{q}} - \frac{\nabla^2}{M_{\s{q}}} + V(r) \right) C_{\s{>}}^{\scriptscriptstyle{PS}}(r,t) = i \frac{\partial C_{\scriptscriptstyle{>}}^{\scriptscriptstyle{PS}}(r,t)}{\partial t},
\label{schdn}
\end{equation}

with the initial condition

\begin{equation*}
C_{\s{>}}^{\s{PS}}(r,0) = -2 M_{\s{q}}^2 N_c \delta^3(\vec{r}).
\end{equation*}

Here $C_{\s{>}}^{\s{PS}}(r,t)$ is real time point-split correlator in pseudo-scalar channel.  

Here, \( V(r) \) is the potential, given by the long-time limit of the Wilson loop \( W_{\s{M}}(r,t) \) as

\begin{equation}
V(r) = i \lim_{t \to \infty} \frac{\partial \log W_{\s{M}}(r,t)}{\partial t}.
\label{pot_def}
\end{equation}

The spectral function for the pseudo-scalar channel is then related to \( C_{\s{>}}^{\s{PS}}(r,t) \) through a Fourier transformation:

\begin{equation}
\rho_{\s{\text{PS}}}(\omega) = \frac{1}{2} \lim_{r \to 0} \int_{-\infty}^{\infty} C_{>}^{\s{PS}}(r,t) \exp(i \omega t) \, dt.
\end{equation}

 As mentioned earlier, in this paper, we will compute the potential in \Cref{pot_def} non-perturbatively on the lattice. We will return to the formulation of this potential calculation on lattice at the end of this section.

Now in the regime $\omega \ll 2\,M_q$ the spectral contributions depend on the quantum number channel of the current. For example, it is well known that in the vector channel, one would expect a diffusion contribution in the $\omega \sim 0$ region, as the vector current is conserved. The diffusion coefficient has been studied extensively on the lattice \cite{Banerjee:2011ra, Banerjee:2022gen, Banerjee:2022uge, Altenkort:2020fgs, Brambilla:2022xbd, Pandey:2023dzz, Sommer:2014mea, Altenkort:2023oms}.   On the other hand, for the pseudoscalar channel, one would not expect a transport peak near the $\omega \sim 0$ region due to the absence of a conservation law for the pseudoscalar channel. It has been shown up to next-to-leading order (NLO) that the spectral function indeed does not contain any transport contribution \cite{Burnier:2017bod}. Therefore, we do not consider any transport contribution in the pseudoscalar channel. 

To obtain the spectral function in the region \(\omega \sim 2 M_{\s{q}}\), we need to compute the potential defined in \Cref{pot_def}, which is related to the real-time Wilson loop. However, since lattice calculations are performed in Euclidean time, an analytic continuation is necessary to extract the real-time Wilson loop. This continuation is achieved through the spectral function of the Wilson loop, as discussed in \cite{Rothkopf:2011db},

\begin{align}
W_{\s{E}}(r,\tau)=\int_{-\infty}^{\infty}\,d\omega\, \rho_{\s{W}}(\omega ,r) \exp(-\omega\,\tau) \\
W_{\s{M}}(r,t)=\int_{-\infty}^{\infty}\,d\omega\, \rho_{\s{W}}(\omega ,r) \exp(-i\,\omega\,t).
\label{Wcorr_sp}
\end{align}

In this paper following \cite{Burnier:2014ssa}, we will calculate the potential from the Coulomb gauge fixed Wilson line correlator.

\section{Lattice Details}
\label{Lattice}
\begin{figure*}
    \centering
    \includegraphics[width=8cm]{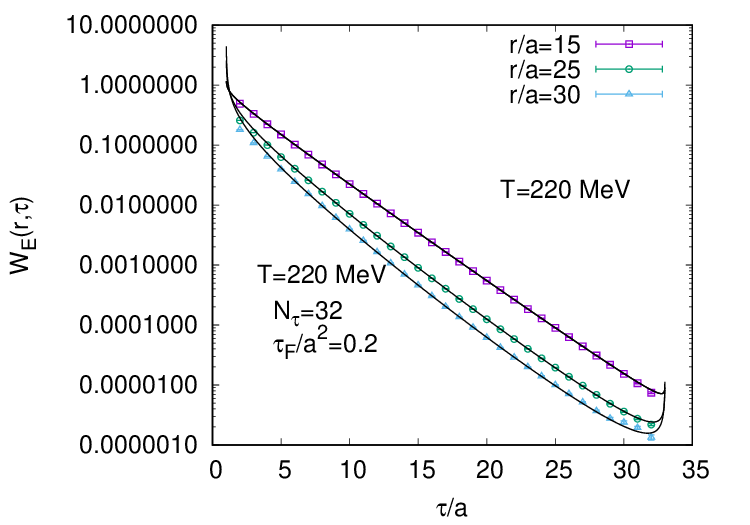}
    \includegraphics[width=8cm]{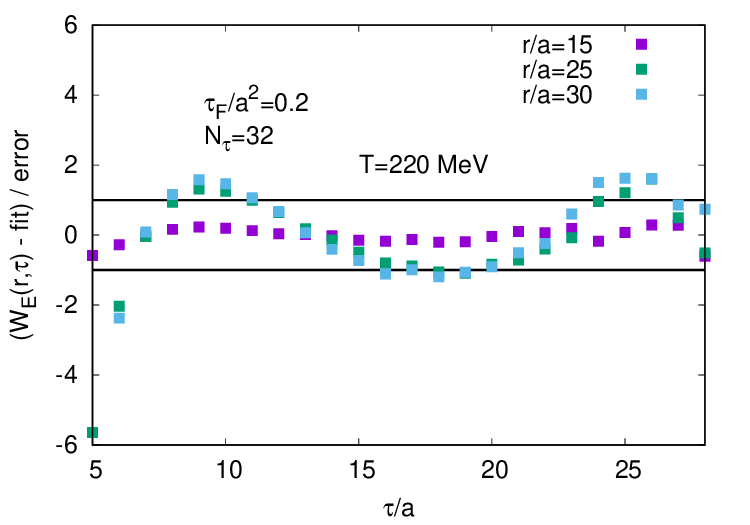}
	\caption{
     Left: The Wilson line correlator from the lattice at T=220 MeV ($N_{\tau}=32$) fitted with \Cref{Param_WL}, excluding the contributions from $c_0$ and $c_1$, for distances $r/a=15, 25,$ and $30$. Right: The corresponding relative error from the fit is shown.
	}
    \label{Fit}
\end{figure*}

\begin{figure*}
    \centering
    \includegraphics[width=8cm]{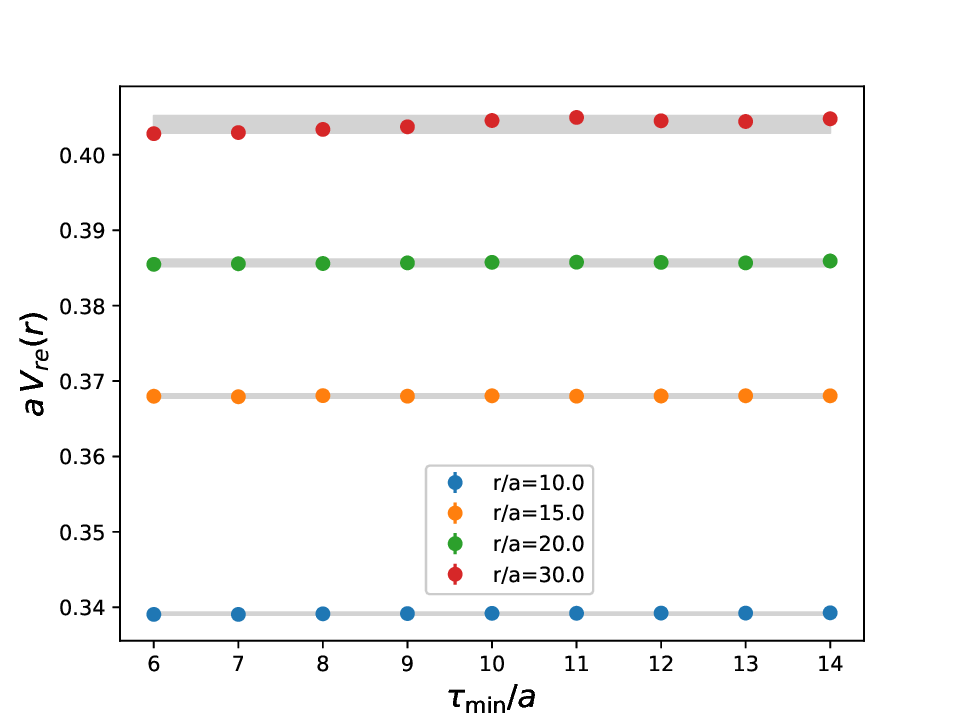}
    \includegraphics[width=8cm]{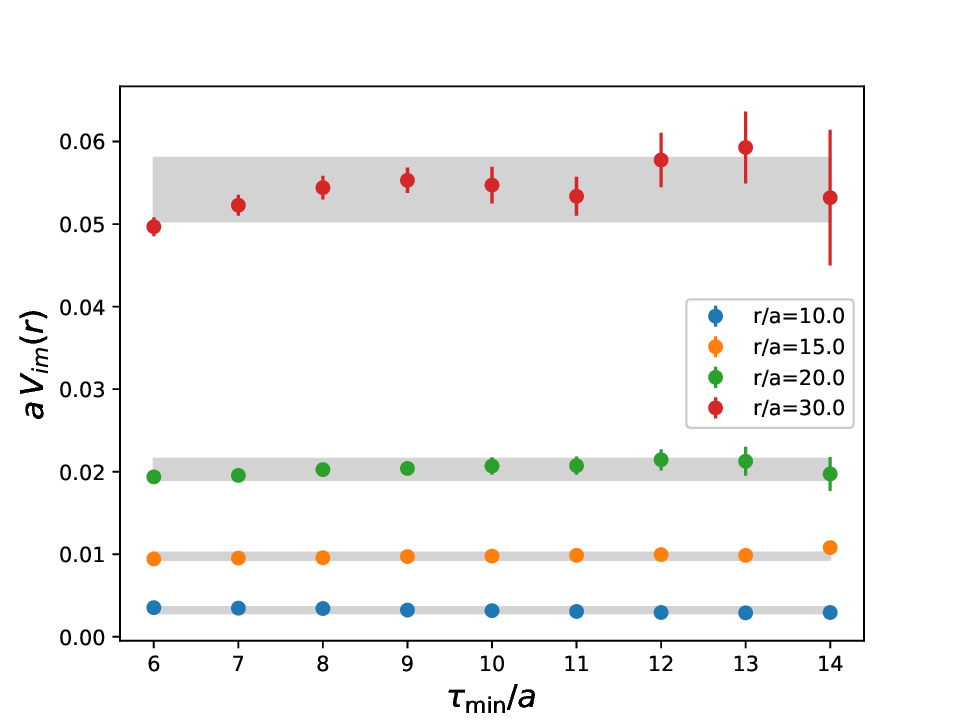}
	\caption{The real part ($V_{re}(r)$) and the imaginary part ($V_{im}(r)$) are plotted as functions of the fit range lower limit $\tau_{min}$. The range has been varied symmetrically with respect to $N_{\tau}/2$, i.e., from $\tau_{min}$ to $N_{\tau} - \tau_{min}$.
    Left: Stability of the real part for $\tau_F/a^2=0.2$. Right: Stability of the imaginary part for $\tau_F/a^2=0.2$. The shaded band represents the value obtained by combining $V_{re}$ and $V_{im}$ from different fit ranges.
	}
    \label{stability}
\end{figure*}

\begin{table}
\centering
\begin{tabular}{c@{\hspace{0.5em}}c@{\hspace{1em}}c@{\hspace{1em}}c@{\hspace{1em}}c@{\hspace{1em}}c@{\hspace{1em}}c}
\hline \\[-1.5mm]
$N_f$ & $\beta_0$  &a [fm]& $T$[MeV] &$N_\sigma^3\times N_\tau$ & confs \\[1.5mm] 
\hline
 \\[-2.5mm] 
\multirow{4}{*}{$2+1$} & \multirow{4}{*}{$8.249$} &\multirow{4}{*}{0.028}& $110$ & $\,64^3 \times 64$ & $1010$ &\\[0.5mm] 
& & & $220$ & $\ 96^3\times 32$ & $1750$  \\[0.5mm]  
& & & $251$ & $\ 96^3 \times 28$ & $612$  \\[0.5mm] 
& & & $293$ & $\ 96^3 \times 24$ & $856 $  \\[0.5mm] 
\hline
\end{tabular}
\caption{
The details of the lattices studied in this paper. The term confs represents the number of configurations used for the computation of the $\eta_c$ and $\eta_b$ correlators } 
\label{tab:lat-detail1}
\end{table}

We calculate the Euclidean correlator of the pseudoscalar $G^{\s{E}}_{\s{PS}}(\tau)$ defined in \Cref{def.pt} for the charm and bottom sectors. To calculate these correlators, we used clover-improved Wilson fermions. The dynamical gauge configurations were generated using a (2+1)-flavor Highly Improved Staggered Quark (HISQ) action for the fermion sector \cite{Follana:2006rc}, and a tree-level improved Lüscher–Weisz gauge action \cite{Luscher:1984xn}, using the RHMC algorithm. The pion mass of these configurations is $m_{\pi} = 320 \, \text{MeV} $, which corresponds to a pseudocritical temperature of $ T_{pc} \sim 180 \, \text{MeV} $. The scale is set by the $f_{\s{K}} = \frac{156.1}{\sqrt{2}} \, \text{MeV} $ \cite{Sommer:2014mea}, correspond to a lattice spacing of $\sim 0.028 \, \text{fm}$. These configurations have also been used in \cite{Altenkort:2023oms, Altenkort:2023eav, Ali:2024xae}.

The parameters of the Wilson clover fermions also need to be fixed. The $c_{\s{SW}}$ coefficient has been fixed to the tadpole-improved value $1/u_0^3$, where $u_0$ is the fourth root of the average plaquette. The hopping parameters $\kappa$ have been tuned such that, at $T = 110 \text{ MeV}$ (zero temperature), the spin-averaged mass $\frac{1}{4}(M_{\s{PS}} + 3M_{\s{V}})$ becomes identical to the PDG values \cite{ParticleDataGroup:2024cfk}. This gives the following $\kappa$ and masses:
\begin{itemize}
    \item For the charmonium sector,  with  $ \kappa_c = 0.13164$,  we obtain $ M_{\eta_{c}} = 2.96 \pm 0.011 \, \text{GeV} $.
    \item For the bottomonium sector, with  $\kappa_b = 0.11684$, we obtain  $ M_{\eta_{b}} = 9.40 \pm 0.02 \, \text{GeV} $.
\end{itemize}

For these quark masses we have calculated the pseudo-scalar correlators \Cref{def.pt} on the lattice at temperatures of \( 220 \, \text{MeV} \), \( 251 \, \text{MeV} \), and \( 293\, \text{MeV} \) in the plasma phase. The lattice spacing is fixed at \(  0.028\, \text{fm} \), and the temperature is varied by adjusting the total temporal extent \( N_{\tau} \). All the lattice details are summarized in \Cref{tab:lat-detail1}. 

We also need to compute the Wilson loop, defined in \Cref{Wcorr_sp}, for the calculation of the potential. It has been shown perturbatively that the spectral function of the Wilson loop contains a significant UV component, in addition to a peak corresponding to the thermal potential \cite{Burnier:2013fca}. This large UV part complicates extraction of thermal peak from the Wilson loop. On the other hand, perturbative calculations also indicate that both the Wilson loop and the Wilson line correlator yield the same thermal potential; however, the UV part is significantly suppressed in the Wilson line correlator \cite{Burnier:2013fca}. As a result, identifying the peak in the spectral function becomes easier in the case of the Wilson line correlator. Consequently, it has become common practice to extract the thermal potential from the Wilson line correlator on the lattice. Following this approach, we calculate the Wilson line correlator fixed in Coulomb gauge.
\begin{table}
\centering
\begin{tabular}{c@{\hspace{0.5em}}c@{\hspace{1em}}c@{\hspace{1em}}c@{\hspace{1em}}c@{\hspace{1em}}c@{\hspace{1em}}c}
\hline \\[-1.5mm]
&$T$[MeV] & $ {\tau_F/a^2}$ & confs& \\[1.5mm] 
\hline
 \\[-2.5mm] 
& $110$ & 0.0, 0.125, 0.2, 0.4 & $1599$ &\\[0.5mm] 
& $220$ & 0.0, 0.125, 0.2, 0.4 ,0.6, 0.8 & $2297$  \\[0.5mm]  
& $251$ &  0.0, 0.125, 0.2, 0.4, 0.6, 0.8 & $2000$  \\[0.5mm] 
& $293$ &  0.0, 0.125, 0.2, 0.4, 0.6, 0.8 & $1380$  \\[0.5mm] 
\hline
\end{tabular}
\caption{Lattices used to measure Wilson line correlators at different flow times and temperatures.} 
\label{tab:lat-detail1_wl}
\end{table}
However, these are purely gluonic observables and we used a much larger-statistics simulation compared to those used for measuring the heavy quark correlators. We also used gradient flow to enhance the signal at large separations. Specifically, we first perform the gradient flow, and then fix the configuration in Coulomb gauge to an accuracy of $10^{-6}$. The details of the configurations used and the flow times are given in  \Cref{tab:lat-detail1_wl}.

\vspace{0.75cm}
\section{Static Thermal Potential}
\label{potential}
\begin{figure*}
    \centering
    \includegraphics[width=8cm]{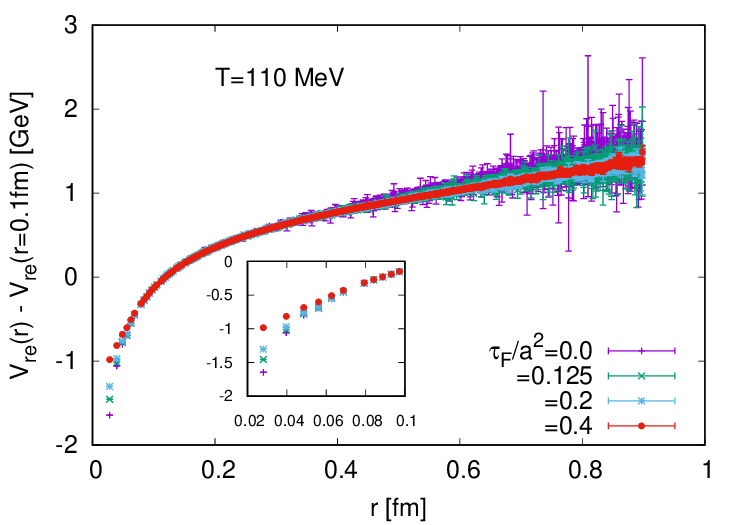}
    \includegraphics[width=8cm]{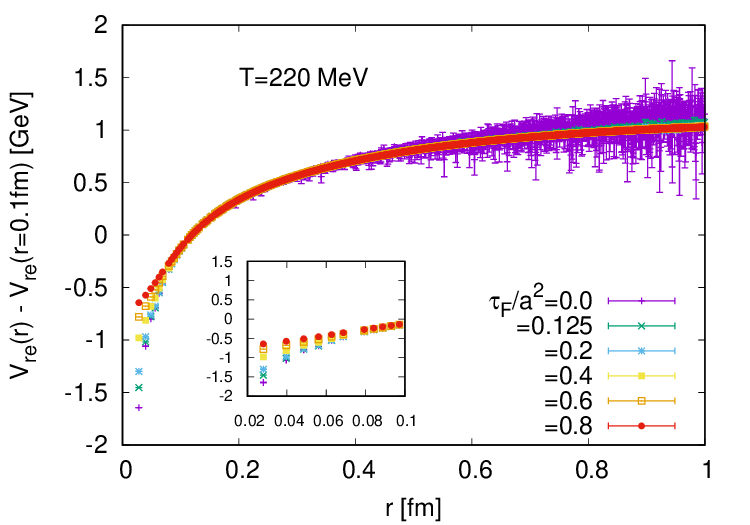}
	\caption{
The flow-time dependence of the real part of the potential at $T = 110$ MeV (left) and $T = 220$ MeV (right).
	}
    \label{Flow_depth}
\end{figure*}

In this section, we discuss the calculation of the static thermal potential non-perturbatively from lattice QCD. The potential, as defined in \Cref{pot_def}, is determined by the long-time behavior of the real-time Wilson loop. At zero temperature, the transfer matrix argument shows that the Euclidean Wilson loop is a sum of exponentially decaying terms, each associated with a different energy state \cite{Philipsen:2013ysa}. At large Euclidean time, the lowest energy state dominates, resulting in a pure exponential decay. After analytic continuation to real time, this gives the static potential in \Cref{pot_def}. This is equivalent to saying the spectral function for the Wilson line correlator in \Cref{Wcorr_sp}, has a well separated lowest lying Dirac Delta peak. 

At finite temperature, the extraction of the potential becomes non-trivial as there is no pure exponential decay. This is because, at finite temperature, the sharp Dirac delta peak acquires a thermal width due to interactions with the plasma, making the extraction of the thermal potential from the Euclidean correlator more difficult. 

In this paper, following \cite{Bala:2019cqu,Bala:2020tdt}, we will use the HTL motivated structure of the correlator and hence the spectral function for the extraction of the potential. Here we will briefly describe the idea behind that. 

The constraints that we would like to put on  the structure of Euclidean Wilson line correlator, is that after the analytic continuation  $\tau\to i\,t$ the limit in \Cref{pot_def} must exist, as the static potential is defined in this limit. 
Since the HTL perturbative calculation of the Euclidean Wilson line correlator in \cite{Laine:2006ns, Burnier:2013fca} shows that this limit exists, we observe the $\tau$ dependence of the HTL perturbative correlator, which leads to the limit \Cref{pot_def} after the analytic continuation $\tau \rightarrow it$. This calculation has been extended to finite flow time in \Cref{appa}. A close observation of this expression shows that the logarithm of the Wilson line correlator can be decomposed into a linear and a periodic part in $\tau$. The linear part of the correlator gives the real part of the potential, as expected, since only the real part of the potential can lead to a pure exponential decay. On the other hand, the periodic part gives rise to the imaginary part and causes the correlator to deviate from exponential decay at large $\tau$. Motivated by this observation, we parametrize the Wilson correlator at large $\tau$ as follows:

\begin{widetext}
\begin{equation}
\log W_{\s{E}}(r, \tau) = -V_{re}(r) \tau + \int_{-\infty}^{\infty} \sigma(r, \omega) \left( \exp(\omega \tau) + \exp(\omega (\beta - \tau)) \right) \, d\omega. \\
\label{W_basic}
\end{equation} 
From this form we can obtain,
\begin{equation}
\lim_{t\rightarrow\infty}i\frac{\partial W_{\s{E}}(r,\tau\to it)}{\partial t}=V_{re}(r)-\lim _{t\rightarrow\infty}\int_{-\infty}^{\infty} d\omega \,\omega \sigma(r,\omega)[\exp(i\omega t)-\exp(\omega(\beta-i t))].   
\end{equation}

Now using the fact that,
\begin{equation}
\lim_{t\rightarrow\infty}\frac{ \exp(i\,\omega\,t) - \exp(\omega (\beta - i\,t))}{\omega}=2\,\pi\,i\,\delta(\omega).
\end{equation}
one can easily see that a finite result is only possible when  $\sigma(r,\omega)\sim \frac{1}{\omega^2}$. In fact this $1/\omega^2$ behaviour can also be seen in HTL expression, where one of the factors comes from the Bose distribution function $n_{b}(\omega)$. This motivates us to write,
\begin{equation}
\sigma(r,\omega)=n_{b}(\omega)\left(\frac{\beta V_{im}(r)}{2\pi\, \omega}+\sum_{l=0}^{\infty} c_{2l+1}\,\omega^{2l+1} \right)    
\end{equation}
Only odd terms are present because the combination $n_{b}(\omega)\left( \exp(\omega \tau) + \exp(\omega (\beta - \tau)) \right)$ is odd in $\omega$.
Using this one gets the following compact form for the Wilson line correlator,
\begin{align}
W(r, \tau) &= A(r)\,\exp\left(-V_{re}(r) \tau -  \frac{\beta\,V_{im}(r)}{\pi }\log\left(\sin\left(\frac{\pi\tau}{\beta}\right)\right)+ \sum_{l=0}^{\infty}\frac{2\,c_{2l+1}}{\beta^{2l+2}} (2l-1)! \left(\zeta\left(2\,l+2,\frac{\tau}{\beta}\right)+\zeta\left(2\,l+2,1-\frac{\tau}{\beta}\right)\right)\right)
\label{Param_WL}
\end{align}
\end{widetext}
Here, $\zeta(p,x)=\sum_{n}\frac{1}{(n+x)^p}$ is generalized zeta function. This parametrization is our ansatz for the calculation of the thermal potential from the lattice data.
This parametrization indeed gives a well defined potential in the long time limit, as can be seen from \Cref{pot_def} after the analytic continuation,
\begin{equation}
V(r)=i \lim_{t\to\infty} \frac{\partial \log W(r,\tau\rightarrow i\,t)}{\partial t}=V_{re}(r)-i\, V_{im}(r).    
\end{equation}
The other terms $c_{1},c_{2},...$ do not survive in the long time limit and hence do not contribute to the static potential. 
\begin{figure*}
    \centering
    \includegraphics[width=8cm]{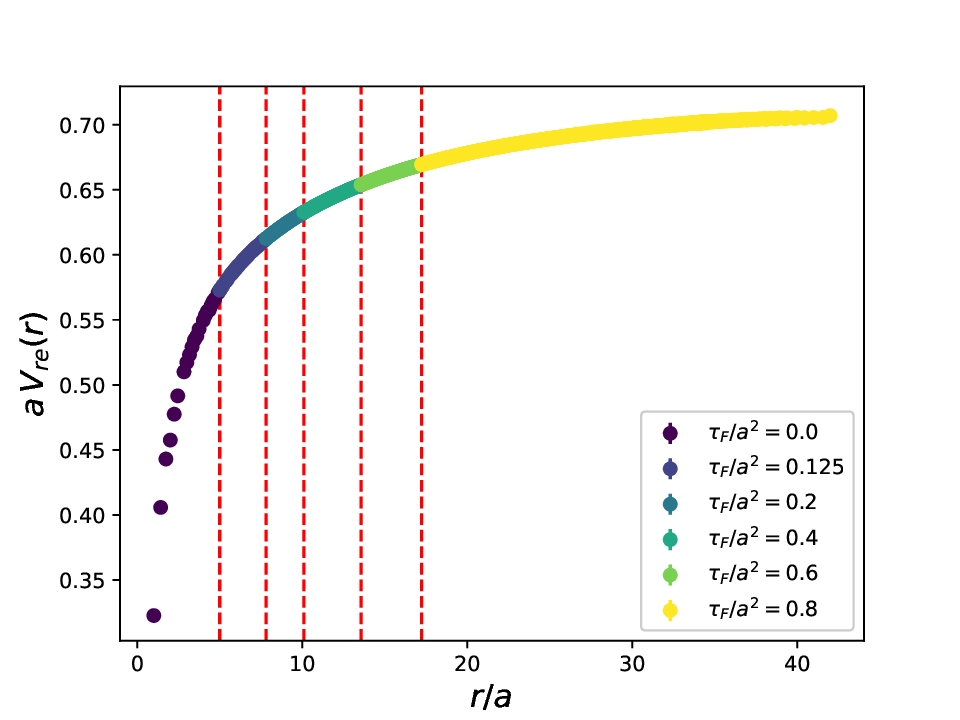}
    \includegraphics[width=8cm]{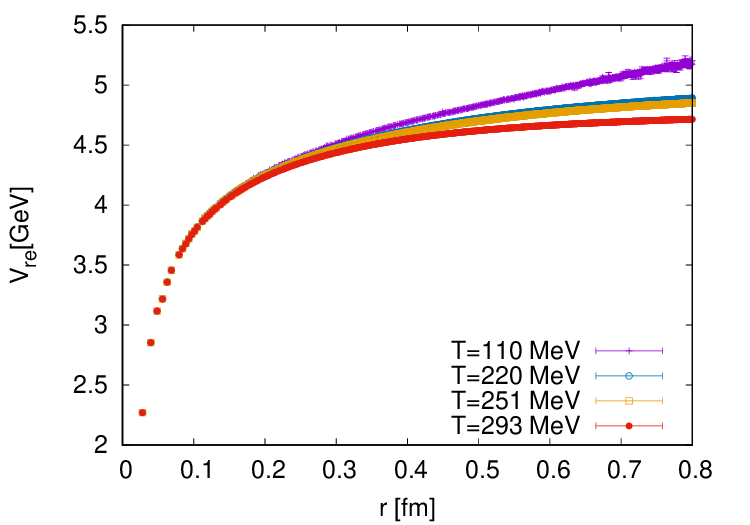}
	\caption{
(Left): The real part of the thermal potential is smoothly matched across different flow times at a temperature of 220 MeV. (Right): Real part of thermal potential at different temperature.}
    \label{Vre_temp}
\end{figure*}

\begin{figure*}
    \centering
    \includegraphics[width=8cm]{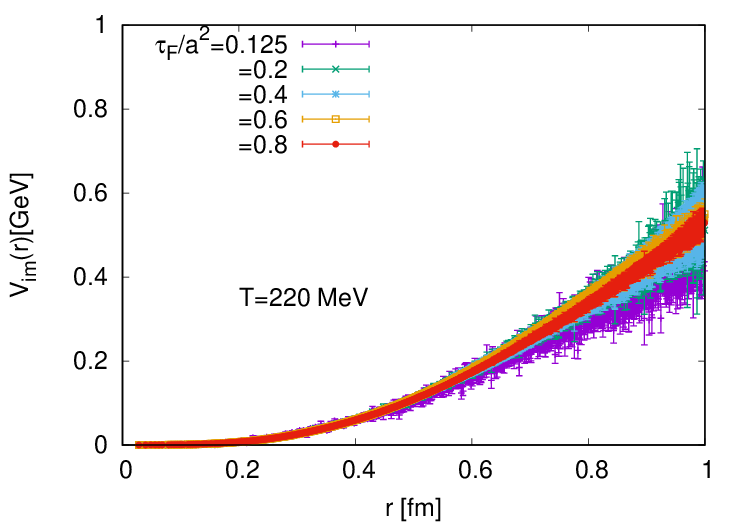}
    \includegraphics[width=8cm]{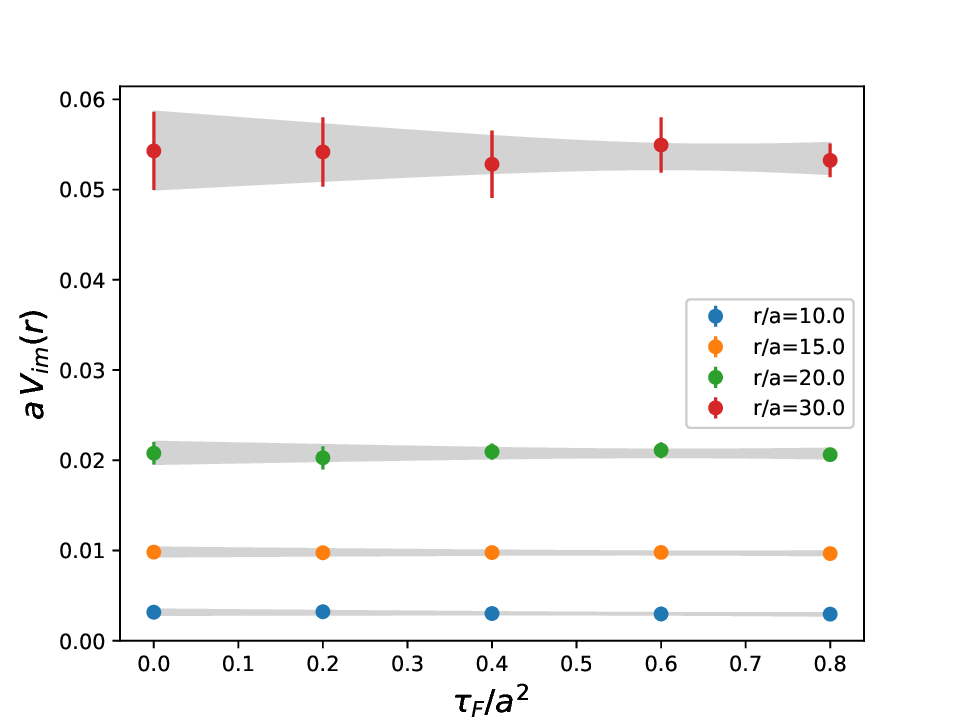}
	\caption{
(Left): Imaginary part at different flow times for $T = 220$ MeV. (Right): Zero flow-time extrapolation at selected distances.
	}
    \label{IM_Flow_depth}
\end{figure*}

We use the parametrization in \Cref{Param_WL} to fit Wilson line correlator at zero as well as at finite flow time. A leading order perturbative analysis of the gradient flow dependence of the Wilson line correlator, as shown in \Cref{appa}, also suggests the structure of \Cref{W_basic}. 
The fit results of the correlator with respect to \( A(r) \), \( V_{re}(r) \), and \( V_{im}(r) \) are shown in the left panel of \Cref{Fit}  for \( N_{\tau} = 32 \) for a few distances at a flow time of \( \tau_{\s{F}}/a^2 = 0.2 \) in the range \( \tau/a = 6-26 \). The right panel shows the relative deviation from the prediction, illustrating predictability. The statistical error on the potential has been determined by fitting our ansatz to each bootstrap sample and taking the standard deviation over them. To find the stability of the fit we also vary the fit range symmetrically around the midpoint, $N_{\tau}/2$. The results for $V_{re}(r)$ and $V_{im}(r)$, shown in \Cref{stability} plotted as a function of the fitting range, show a very mild dependence on the fitting range, illustrating the stability of the fit.
The statistical average over the fitting range, shown by the shaded band in the figure, is taken to be the final value of the potential from a given flow depth. The real part of the potential for various flow depths at temperatures of 110 MeV and 220 MeV is shown in \Cref{stability}. The temperature 110 MeV is much smaller than the crossover temperature and the potential at this temperature is purely real. Therefore the potential for this temperature has been extracted using a standard exponential fit to the Wilson line correlator at large $\tau$. As the real part part of the potential contains an additive renormalization, which depends on the flow depth, we have plotted $V(r)-V(r=0.1\, \text{fm})$ in  \Cref{Flow_depth}. As one can see, the flow reduces the noise of the potential at large distances. But at same time the flow deforms the short distance part of the potential. This behaviour can also be observed in perturbative calculation of the flowed potential in \Cref{FT_LO} of \Cref{appa}. Therefore, we calculate the potential at several flow times, including zero flow depth. The real part of the potential at short distances is taken from the zero flow time correlator and it is smoothly connected to the larger flow depth potential as shown in the left panel of \Cref{Vre_temp}. On the right panel of \Cref{Vre_temp}, we plot the real part of the potential for various temperatures. As we can see, below \( T_{pc} \) the potential is confined, but as the temperature rises above \( T_{pc} \), the potential shows a much flatter behaviour because of the medium modification due to the presence of color screening in the QGP phase, and the strength of screening increases with temperature. This result is in agreement with earlier studies in \cite{Burnier:2014ssa, Bala:2019cqu, Burnier:2015tda}. However, it contradicts the study in \cite{Bazavov:2023dci}, where no color screening is observed up to a temperature of $\sim 350\, \text{MeV}$. The contradiction serves as an indication of the inversion problem’s ill-posed nature in \Cref{Wcorr_sp} and the physics input required for meaningful spectral reconstruction.  
 The short distance part of the potential in \Cref{Vre_temp} is from zero flow time correlator, however this region also gets affected by the lattice artifacts and in the next section we will replace this region by 3-loop perturbative potential. 

\begin{figure*}
    \centering
    \includegraphics[width=8cm]{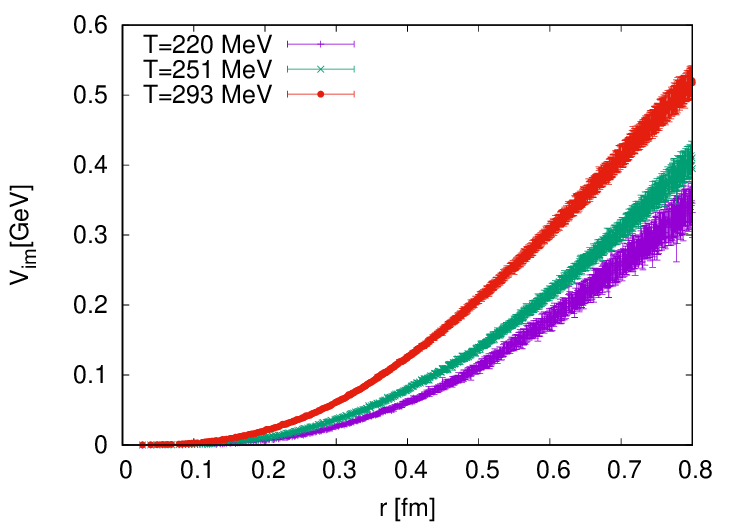}
    \includegraphics[width=8cm]{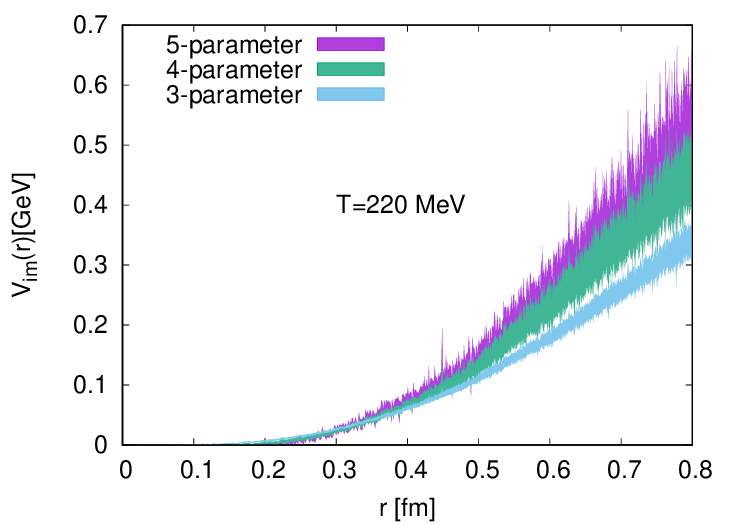}
	\caption{
(Left): Zero flow-time extrapolated imaginary part at different temperatures.
(Right): Dependence of the zero flow-time extrapolated imaginary part on the fit parameters.
	}
    \label{Vim_temp}
\end{figure*}

The imaginary part for different flow time has been shown in \Cref{IM_Flow_depth} for temperature 220 MeV. Though the perturbative prediction in \Cref{appa} shows that imaginary part gets affected by the gradient flow, we do not find much flow time dependence in the lattice data within error. Nevertheless we have performed a linear extrapolation to zero flow time as shown in \Cref{IM_Flow_depth}.  
The zero flow time extrapolated imaginary part for different temperature is shown on the right panel of \Cref{Vim_temp}. The imaginary part is increasing with both distances and temperature. The imaginary part is expected to saturate at large distances because, at large distances, the quark-antiquark pair is expected to be scattered by thermal gluons independently. However, in the distance scale we studied on the lattice, we do not see any such saturation.

In this section, we have shown the results using the first three parameters in \Cref{Param_WL}, namely \( A(r) \), \( V_{re}(r) \), and \( V_{im}(r) \), which already describe a large range of the Wilson line correlator. We have also performed the fit, including \( c_1 \) and \( c_2 \) individually. We find that although there is no change in the real part of the potential, the imaginary part of the potential shows slight changes at large distances, and the error bars increase, as shown in the right panel of \Cref{Vim_temp}. We have taken this spread in the imaginary part as a systematic uncertainty in determining the imaginary part, and it has been included in the determination of the quarkonia spectral function in \Cref{Spec fn}.

\section{Pole-Mass extraction and Parameterization of thermal potential}
\label{Pole_mas}
We need to obtain the quark mass appearing in the Schr\"odinger Equation for spectral function in \Cref{schdn}, which we will call ``pole-mass" in our context. We will use the zero-temperature potential to extract the pole mass of heavy quarks. However the potential calculated from lattice is defined up to an additive renormalization constant. On the other hand the short distance part of the potential obtained from lattice has artifacts from finite lattice spacing. The short distance ($r\ll \frac{1}{\Lambda_{QCD}}$) part of the potential however can be calculated reliably from perturbation theory. Since the potential is temperature independent in this region, one can use vacuum perturbation theory. The perturbative potential up-to 3-loop is given by \cite{Smirnov:2009fh, Sumino:2014qpa},
\begin{widetext}
\begin{align}
V(r) &= \int \frac{d^3 \vec{q}}{(2\pi)^3} \exp(i\vec{q} \cdot \vec{r}) \frac{\alpha_{\s{V}}(q^2)}{q^2} \nonumber \\
\label{3-loop}
\text{where,}\nonumber\\
\alpha_{\s{V}}(q^2) &= - 4 \pi C_{\s{F}} \alpha_s (|\vec{q}|) \left\{ 1 + \frac{\alpha_s(|\vec{q}|)}{4 \pi} a_1 + \left( \frac{\alpha_s(|\vec{q}|)}{4 \pi} \right)^2 a_2 + \left( \frac{\alpha_s(|\vec{q}|)}{4 \pi} \right)^3 a_3  + \mathcal{O}(\alpha_s^4) \right\}
\end{align}
\end{widetext}
Here, $a_1,a_2$ and $a_3$ depend on number of flavor $N_f$ and number of color $N_c$. For the running coupling  $\alpha_{s}= g^2/4 \pi$ 4-loop beta function \cite{Herzog:2017ohr} has been used.

However, the perturbative potential suffers from a well-known renormalon problem originating from the integration region \( |\vec{q}| = \Lambda_{\text{QCD}} \), where the QCD coupling constant has a pole \cite{Sumino:2014qpa}. Therefore, the perturbative potential soon becomes unreliable, and smooth matching with the lattice potential becomes difficult.
 As a result, we subtract the renormalon contribution by avoiding integration in \Cref{3-loop} over the entire region \( 0 < |\vec{q}| < \infty \). Instead, we perform the integration only up to some cutoff \( \mu_r< |\vec{q}| < \infty \), where \( \mu_r \gg \Lambda_{\s{\text{QCD}}} \). This will create an additive renormalization of $O(\Lambda_{\s{\text{QCD}}})$ originating from renormalon pole and weak $r^2$ dependence, which can be neglected at short distance \cite{Sumino:2014qpa}. Nevertheless, renormalon subtraction will make the perturbative potential well behaved up to a larger distance and a smooth matching with lattice QCD potential is possible. We checked that the short distance part of the potential has a weak dependence on the value of $\mu_r$ up to an additive constant and we fix the value $\mu_r=1\,\text{GeV}$.
 
To match the zero temperature lattice potential with this renormalon-subtracted potential, we first fit the lattice potential using the Cornell form:
\begin{equation}
    V(r) = -\frac{\alpha}{r} + \sigma r + c
\end{equation}
within the range \([r_{\text{min}}, r_{\text{max}}]\). We find that setting \( r_{\text{min}} = 3\,a \) provides a good description of the lattice data up to  \( r_{\text{max}} =25\,a\). From this fit we  obtained a string tension $\sqrt{\sigma}= 0.45\,\text{GeV}$ and  $\alpha= 0.42$. After the parametrization, we performed a smooth matching with the RS potential at a distance of \( r_{m} \sim 0.1\,\text{fm} \), by adjusting the additive renormalization of the lattice potential. However, this matching still carries an additive uncertainty of \( \mathcal{O}(\Lambda_{\s{\text{QCD}}}) \) from the renormalon pole in the RS potential, which needs to be fixed.

\begin{figure*}
    \includegraphics[width=8cm]{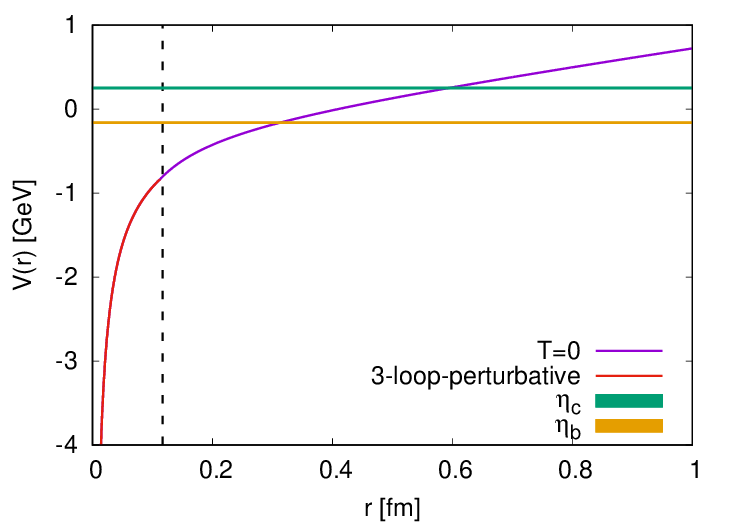}
	\caption{
The lattice zero-temperature potential smoothly matches the renormalon-subtracted vacuum potential at short distances. The overall additive constant is fixed to reproduce the ground state masses of $\eta_b$ and $\eta_c$.}
    \label{Param_real}
\end{figure*}

To fix this additive constant, we use the $\eta_{c}(1S)$ and $\eta_{b}(1S)$ masses extracted from lattice correlators, as described in \Cref{Lattice}. The pole mass of the bottom quark is fixed to $m_b = 4.78\,\text{GeV}$, as quoted by the PDG \cite{ParticleDataGroup:2024cfk}. Using this mass, we solve the Schr\"odinger equation and adjust the additive constant of the potential such that the ground state mass matches the \(\eta_{b}(1S)\) mass obtained from the lattice.  Using the same additive constant, we then tune the charm quark pole mass by solving the Schr\"odinger equation so that the ground state mass matches the \( \eta_c(1S) \) mass obtained from the zero-temperature lattice correlator. This leads to a pole charm quark mass $1.35\pm 0.01\, \text{MeV}$. The matched zero-temperature potential, after fixing the additive constant, is shown in the left panel of \Cref{Param_real}. These mass estimates, of course, suffer from an uncertainty of order \( O(\Lambda_{\text{QCD}}) \), but this uncertainty can always be absorbed into the renormalon ambiguity of the potential, since the combination  $2\,M + V(r)$ is expected to be free from renormalon ambiguity \cite{Sumino:2014qpa}. 

For the finite temperature potential in \Cref{Vre_temp} we see that up to the matching distance ($r_m = 0.1\,\text{fm}$), thermal effects in the potential can be neglected in the temperature range we studied.
In this region, we replace the thermal potential with the perturbative potential, therefore we subtract the same additive constant that has been used at zero temperature. This is crucial to always replace the temperature-independent part of the thermal potential with the zero-temperature potential and using the same additive constant from the zero-temperature matching. This ensures that the relative separation between the zero and finite temperature potentials remains same. 

In the spectral function calculation one has to solve the Schr\"odinger Equation with the thermal potential, for which we need a parametrization of the the thermal potential as well that describe the lattice data in \Cref{Vre_temp}. We parameterize the finite temperature potential with the KMS form \cite{Karsch:1987pv}, 
\begin{equation}
V_{re}(r,T)=-\frac{\alpha(T)}{r}\,\exp(-m_{r} r)+\frac{\sigma}{m_r} (1-\exp(-m_r r))+C.
\label{real_param}
\end{equation}
We fit the thermal lattice potential with this form with respect to the parameters $\alpha,\sigma,m_r,C$. We found a reasonable good description of our lattice data with this form in the range $r_\text{min}=3\,a$ to $r_\text{max}=35\,a$. The fitted parameter are given in \Cref{tab:fit_param_real}. In the table we have shown the value obtained from the fits. We do not quote the error bar, as it is negligible compared to the uncertainty introduced in the spectral function due to mass tuning.

\begin{figure*}
    \centering
     \includegraphics[width=8cm]{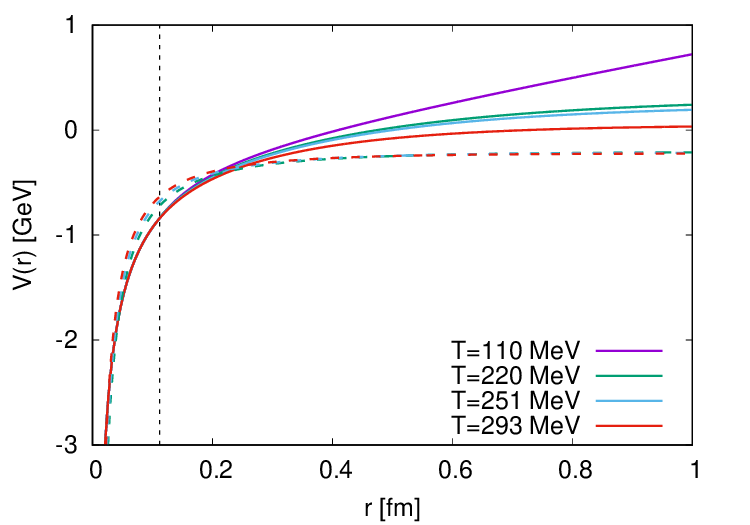}
    \includegraphics[width=8cm]{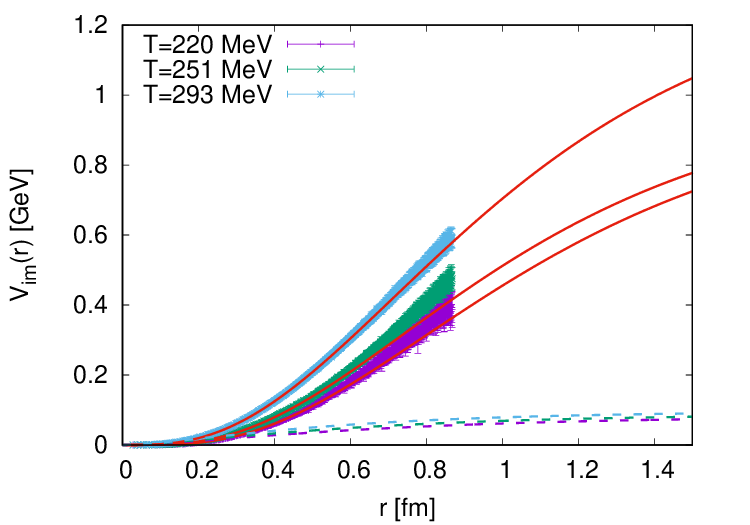}
	\caption{
(Left): The thermal potential matches the renormalon-subtracted vacuum potential at short distances, using the same additive constant as the zero-temperature potential. The perturbative potential is shown with dashed lines.
(Right): Parametrization of the imaginary part.
	}
    \label{Vim_param}
\end{figure*}

\begin{table}
\centering
\begin{tabular}{c@{\hspace{1em}}c@{\hspace{1em}}c@{\hspace{1em}}c@{\hspace{1em}}c}
\hline \\[-1.5mm]
$T$ [MeV] & $\alpha(T)$ & $\sqrt{\sigma(T)}$ [GeV] & $m_r(T)$ [GeV] & $C(T)$ [GeV] \\[1.5mm] 
\hline
 \\[-2.5mm] 
$220$ &$0.326$ &$0.710$  &$0.521$ & $4.047$\\[0.5mm]  
$251$ & $0.314$& $0.746$&$0.583$ & $3.992$\\[0.5mm] 
$293$ &$0.308$ &$0.832$ &$0.803$ & $3.889$\\[0.5mm] 
\hline
\end{tabular}
\caption{Parameters at different temperatures for real part paramerization.} 
\label{tab:fit_param_real}
\end{table}

The matched thermal potential along with the zero temperature potential is shown in \Cref{Param_real}. In the same plot we plotted the perturbative thermal potential from \cite{Laine:2006ns} given by,
\begin{equation}
    V^{re}_{\s{T}}(r) = -\frac{g(T)^2}{4\pi} C_{\s{F}} \biggl[ m_d^{ } + \frac{\exp(-m_d^{ } r)}{r}
 \biggr]\;
\end{equation}
We used the two-loop formula for the thermal coupling and Debye Mass from \cite{Laine:2005ai}. We find that color screening is weaker in the non-perturbative potential than in the perturbative one.

We close the discussion of this section by also providing a parametrization of the imaginary part shown in \Cref{Vim_temp}, which will also be needed in the next section for the spectral function calculation. Motivated from \cite{Guo:2018vwy}, we parameterize the imaginary part by,

\begin{equation}
V_{im}(r)= a_{1} \phi_2(m_{i}r) + a_{2} \phi_3(m_{i}r) + a_{3} \phi_4(m_{i}r),
\label{im_param}
\end{equation}
where
\begin{equation}
\phi_{n}(x) = 2\int_{0}^{\infty} \frac{z \, dz}{(z^2+1)^n} \biggl[1-\frac{\sin(z x)}{z x}\biggr].
\label{imag_param}
\end{equation}
Here $a_1,a_2,a_3$ and $m_i$ are the fit parameters. This parametrization is consistent with the expectation that at large distances, the imaginary part should saturate to a finite value.

The parametrization is illustrated in \Cref{Vim_param} for the case of a number of parameters of three in \Cref{Param_WL}. We have also performed fitting with the above form, using the imaginary part obtained by changing the number of parameters to 4 and 5 in \Cref{Param_WL}. The error in the imaginary part becomes larger at large distances. As a result, we perform the parametrization for various upper limits of the fit range. In \Cref{app_b}, we study the impact of this fit range on the spectral function, which turns out to be very small. Further details of the systematic uncertainty estimation of the spectral function from the imaginary part are given in \Cref{app_b}.

In \Cref{Vim_param} we also show the HTL perturbative imaginary part given by 
\begin{equation}
    V^{\text{im}}_{\s{T}}(r) =  { \frac{g^2}{4\pi} C_{\s{F}} T } \, \int_{0}^{\infty} \frac{z \, dz}{(z^2+1)^2}\biggl[1-\frac{\sin(z m_d r)}{z m_d r}\biggr]\;.
\end{equation}

We observe that the non-perturbative imaginary part is significantly larger than the perturbative one, resulting in a higher dissociation rate of quarkonia in the plasma.

We emphasize that the forms given in \Cref{real_param} and \Cref{im_param} are simple parametrizations used solely for solving the Schr\"odinger equation. Any other parametrization consistent with the lattice data should also be allowed. Therefore, these forms are purely phenomenological in nature.

\begin{figure}
    \centering
    \includegraphics[width=8cm]{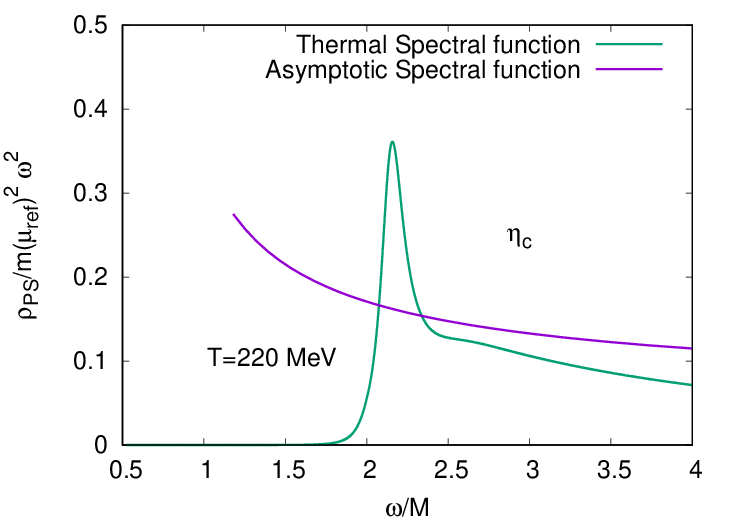}
    \includegraphics[width=8cm]{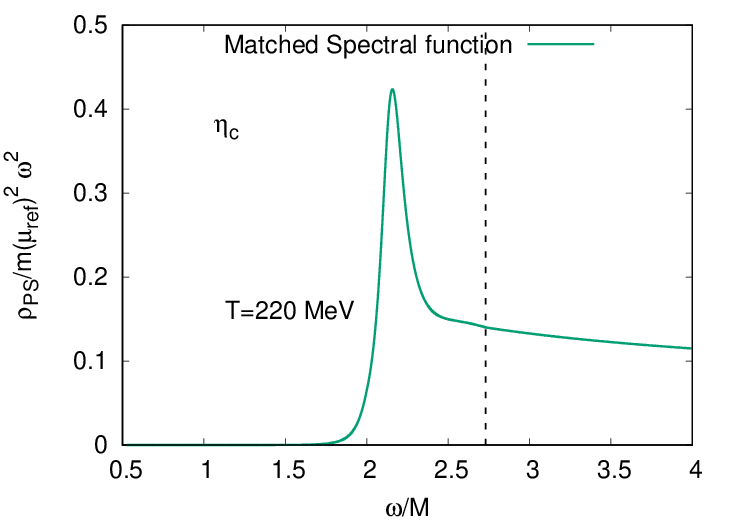}
	\caption{
	Top: The spectral function obtained from the thermal potential in the threshold region at $T=220$ MeV temperature (green curve). The blue curve represents asymptotic vacuum perturbative spectral function in the PS channel.
	Bottom: The matched spectral function in the entire region. 
	}
	\label{matching}
\end{figure}

\section{Spectral function}
\label{Spec fn}
In this section, we will discuss the computation of the spectral function for the correlator given in \Cref{def.pt}. Let's begin with the calculation of the spectral function in the region $\omega \gg 2 M_q $. In this region temperature effects are power suppressed and we use zero temperature perturbation theory.  As discussed in \cite{Burnier:2017bod}, the vacuum perturbative spectral function normalized by the pole quark mass, $\frac{\rho_{\s{\text{PS}}}}{M_q^2 \omega^2}$, shows poor convergence even at large $\omega$, whereas when it is normalized by the $\overline{\text{MS}}$ mass, $\frac{\rho_{\s{\text{PS}}}}{m(\bar \mu)^2 \omega^2}$ shows relatively good convergence at large $\omega$.

We estimate the charm and bottom quark $\overline{\text{{MS}}}$ mass from the corresponding pole quark mas, obtained from the zero-temperature potential in \Cref{Pole_mas}. We use the self-consistency condition from \cite{Burnier:2017bod}, namely at large $\omega$ one would expect, $\frac{\rho_{\s{\text{PS}}}}{M_q^2 \omega^2}=\frac{\rho_{\s{\text{{PS}}}}}{m(\bar \mu)^2 \omega^2}|_{\mu=\omega, \omega=x M_q}$, where we take $x=6$. The charm and bottom quark  $\overline{\text{{MS}}}$ mass obtained in this way is given by $1.04\,\pm\, 0.01\,\text{GeV} $ and $5.3\,\text{GeV}$ at the scale $\mu_{\text{ref}}=2\,\text{GeV}$. Again the errorbar in the charm mass is due to the uncertainty in tuning of zero temperature $\eta_c(1S)$ and $\eta_b(1S)$ mass. 

Near the threshold region $\omega\sim 2\,M_q$, thermal effects become important, and the spectral function needs to be calculated using the thermal potential from \Cref{Vim_param}, following the formalism outlined in \Cref{Formalism}. We solve the Schr\"odinger equation in \Cref{schdn} using the algorithm given in \cite{Burnier:2007qm} to obtain the spectral function in this region. Much below the threshold, $\omega \ll 2\,M_q$, the Schr\"odinger equation overestimates the spectral function. The NLO calculation given in \cite{Burnier:2017bod} shows that in the $\omega < 2\,M_q$ region, the spectral function undergoes exponential suppression. This kind of suppression is not included in the Schr\"odinger estimate of the spectral function and requires further modeling in this region. Motivated by this, and following \cite{Burnier:2017bod}, we multiply the imaginary part in the $\omega < 2\,M_q$ region by $\exp\left(\frac{2\,M_q - \omega}{T}\right)$ to model this suppression. The spectral function in both the asymptotic and thermal regions has been smoothly matched as follows:
\begin{equation}
\rho_{matched}=A_{0}\,\rho_{\s{T}}(\omega) \theta(\omega_{0}-\omega)+\rho_{vac}(\omega)\theta(\omega-\omega_{0}) .
\end{equation}

\begin{figure*}
    \centering
    \includegraphics[width=8cm]{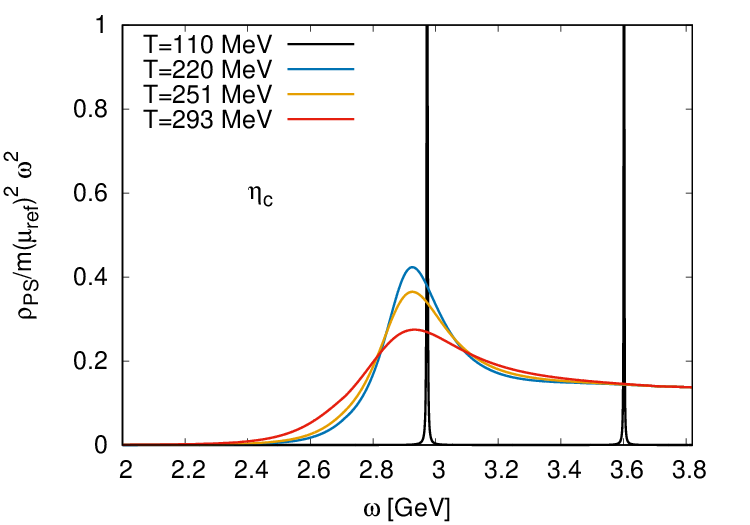}
    \includegraphics[width=8cm]{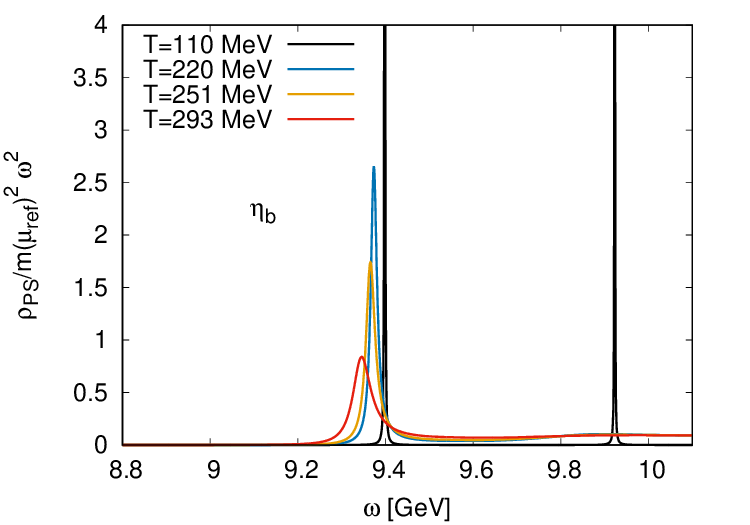}
	\caption{
The spectral function at various temperatures for the $\eta_c$ state is shown on the left, and for the $\eta_b$ state on the right.
	}
\label{sp_temp}
\end{figure*}

\begin{figure*}
    \centering
    \includegraphics[width=8cm]{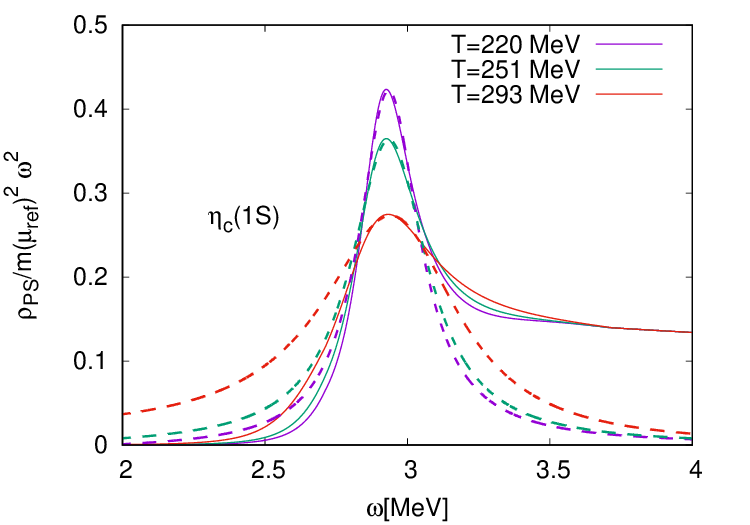}
    \includegraphics[width=8cm]{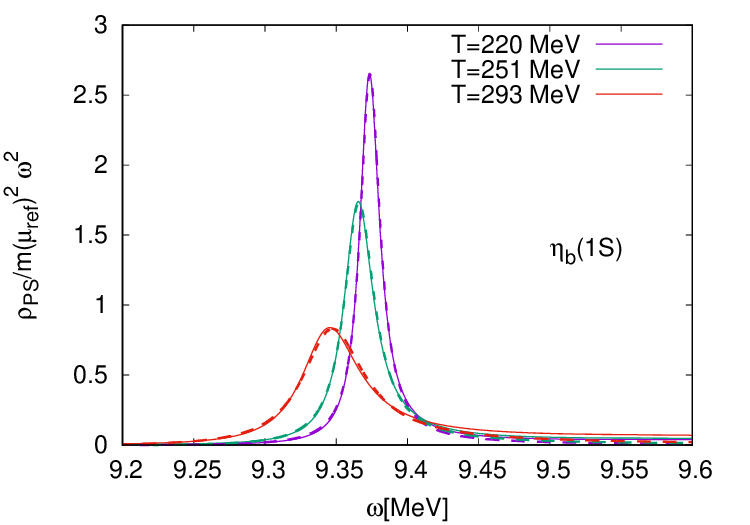}
	\caption{
The spectral function has been fitted near the peak using \Cref{sp_param}. The left panel shows the $\eta_c$ case, while the right panel corresponds to $\eta_b$.
	}
    \label{Param_sp}
\end{figure*}

The parameters \(A_0\) and \(\omega_0\) were determined by equating \(A_0 \rho_{\s{T}}(\omega_0)\) with \(\rho_{vac}(\omega_0)\) at \(\omega_0\), along with a smooth matching at \(\omega_0\). The matching is performed in a region where the thermal spectral function has no temperature dependence, as this region is being replaced with the vacuum spectral function. This keeps the relative effect of temperature on the spectral function near the threshold unchanged after the matching. We find that \(\omega \sim 2.7 M_q\) for charm and \(\omega \sim 2.2 M_q\) for bottom serve this purpose with a smooth matching. We found $A_0\sim 1$ for both cases. 
 The spectral function matching procedure is illustrated in \Cref{matching}. In the top panel of \Cref{matching}, the spectral functions from the thermal potential and from the perturbative vacuum regions are shown separately, while in the bottom panel, they are smoothly matched. For the pseudoscalar channel, we do not need to consider the region for $\omega \sim 0$, as there is no transport contribution in this channel. 

The spectral functions for the $\eta_c$ and $\eta_b$ at various temperatures are shown in \Cref{sp_temp}, with the left and right panels displaying them, respectively. At a temperature of 110 MeV, we observe a bound state peak corresponding to the peak position of the vacuum bound state mass. As the temperature rises above the crossover temperature, we observe significant thermal modification of the spectral functions. We see in both, the $\eta_c$ and $\eta_b$ channel, the excited state peaks melt at the temperatures we studied.  The sharp ground state peak, however, broadens due to thermal interactions and its peak position shifts towards smaller masses depending on the quark mass. This is consistent with the spectral function of the point source from the $T$-matrix approach \cite{Liu:2017qah, Tang:2024dkz}, where no excited state peaks were observed in the temperature range we studied. However, in the study \cite{Tang:2024dkz}, when an extended source was used for bottomonium, excited state peaks did appear—but with very large thermal widths. As noted in that work, the existence of such peaks becomes questionable due to the large width associated with the extended operator. Similarly, when model spectral functions such as a Gaussian or cut-Lorentzian were used to analyze NRQCD bottomonium correlators with extended operators, a large width was also observed for excited state \cite{Larsen:2019bwy,Larsen:2019zqv, Ding:2025fvo}. 

This broadening is attributed to the imaginary part of the potential, which represents the thermal decay width due to the inelastic scattering of heavy quarks by plasma constituents. The real part of the potential leads to a screening effect, shifting the peak position towards lower values of $\omega$. The imaginary part on the other hand tends to push the peak towards higher $\omega$. The interplay between these two effects determines the final position of the peak.

In the left panel of \Cref{sp_temp}, we observe that the $\eta_c(1S)$ state has a much larger thermal decay width compared to the thermal decay width of the $\eta_b(1S)$ state. This difference is expected, given that the charm quark mass is significantly smaller than the bottom quark mass, leading to stronger thermal effects in the charm sector. As the temperature increases, the broadening of these states becomes more pronounced. At approximately 293 MeV, the ground state in the charm sector shows a significant increase in broadening, while in the bottom sector, a well-defined ground state peak still persists at this temperature.

The thermal potential and quark masses used to obtain the spectral function have both systematic and statistical uncertainties. However, the systematic uncertainty is larger compared to the statistical uncertainty. We find that most of the dominant source of error arises from the uncertainty in determining the quark masses from the zero-temperature bound state mass and from the imaginary part of the potential, which depends on the number of parameters used. As a result, we study the spectral function by varying the quark mass and the imaginary part within there errorbar as discussed in \Cref{app_b}. In \Cref{sp_temp}, the spectral function is computed for the central value of the quark masses and the imaginary part obtained from a three-parameter fit. To assess the effect of systematic uncertainty, we study the width of the peak and the mass position as a function of temperature.

\begin{figure*}
    \centering
    \includegraphics[width=8cm]{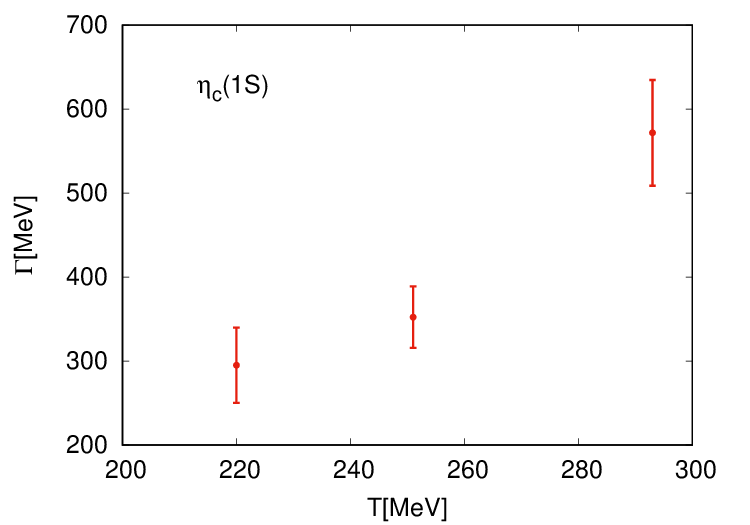}
    \includegraphics[width=8cm]{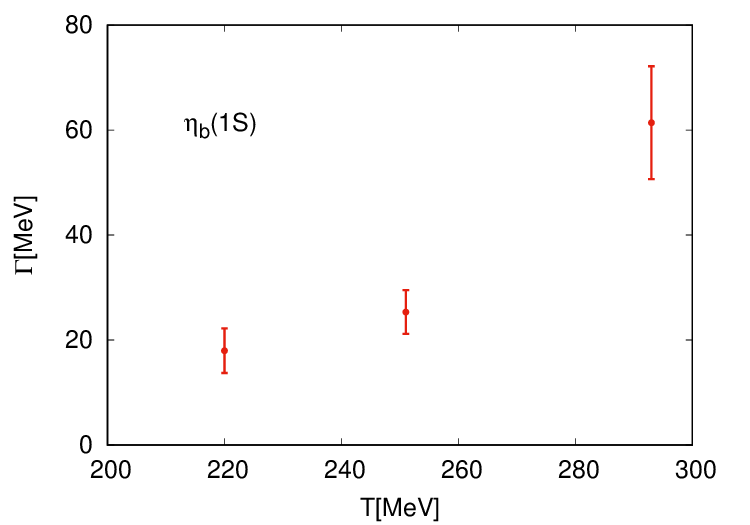}  
    \includegraphics[width=8cm]{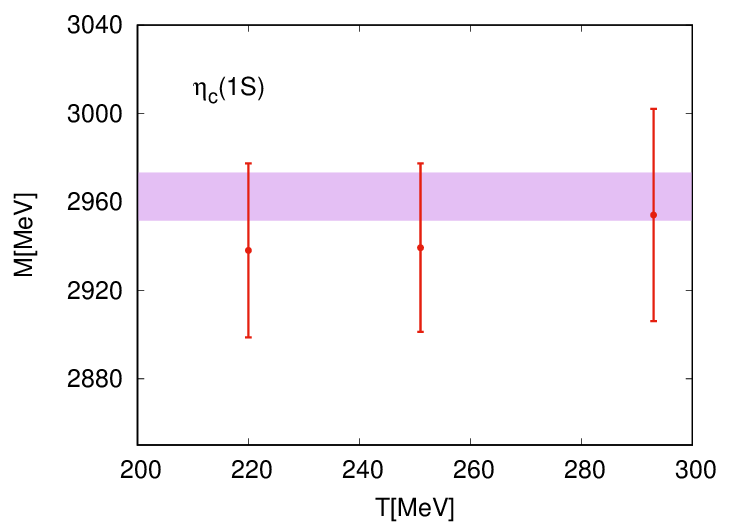}
    \includegraphics[width=8cm]{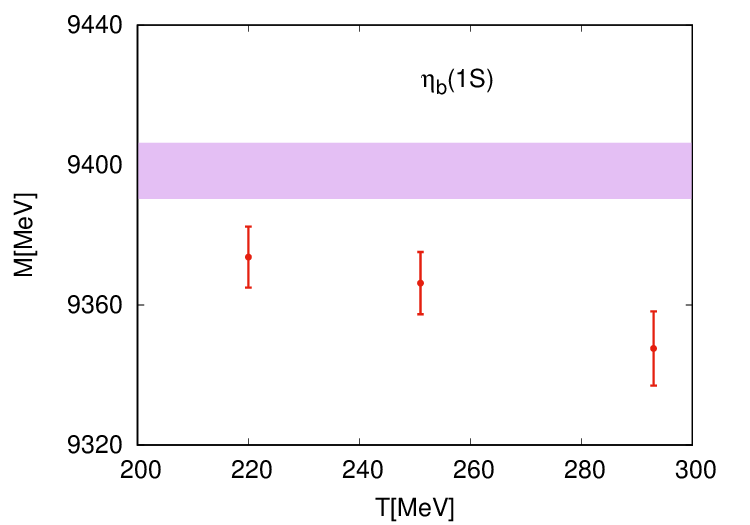}
	\caption{
Thermal mass shifts and decay widths of the quarkonium $1S$ states are extracted by fitting the spectral function near the peak, as described in \Cref{sp_param}.
Top panels: Decay width as a function of temperature for the charm sector (left) and bottom sector (right).
Bottom panels: Thermal mass as a function of temperature for the charm sector (left) and bottom sector (right).
	}
    \label{width}
\end{figure*}

\begin{figure*}
    \centering
    \includegraphics[width=8cm]{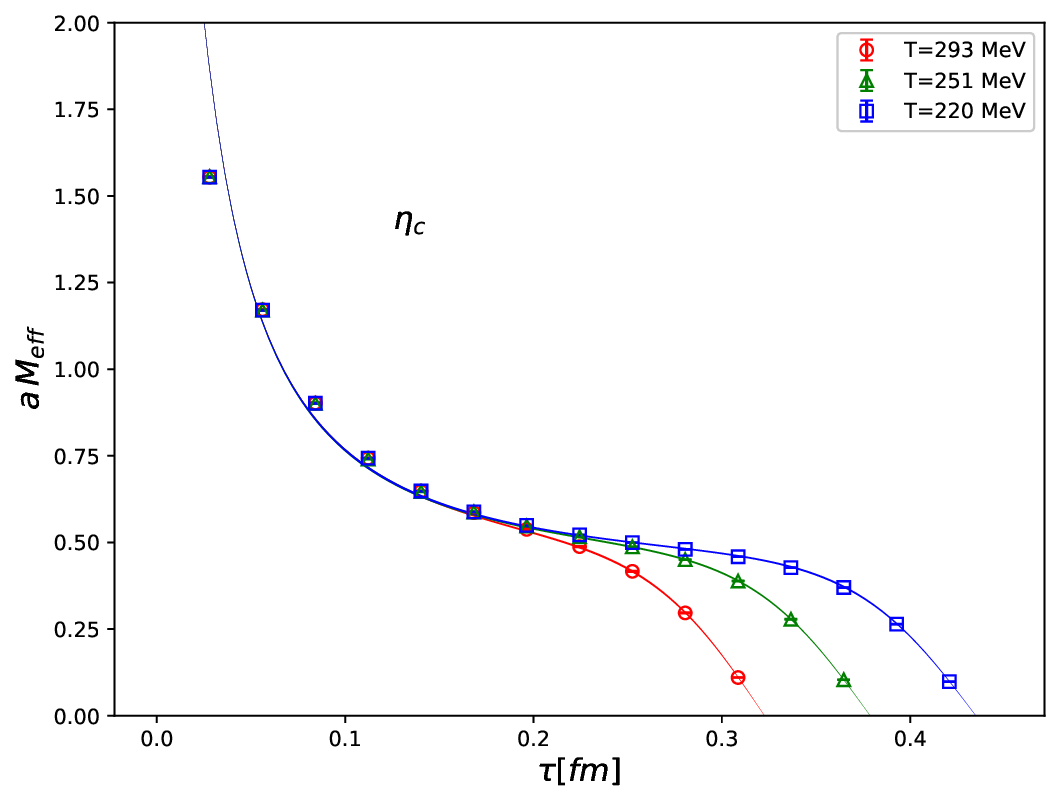}
    \includegraphics[width=8cm]{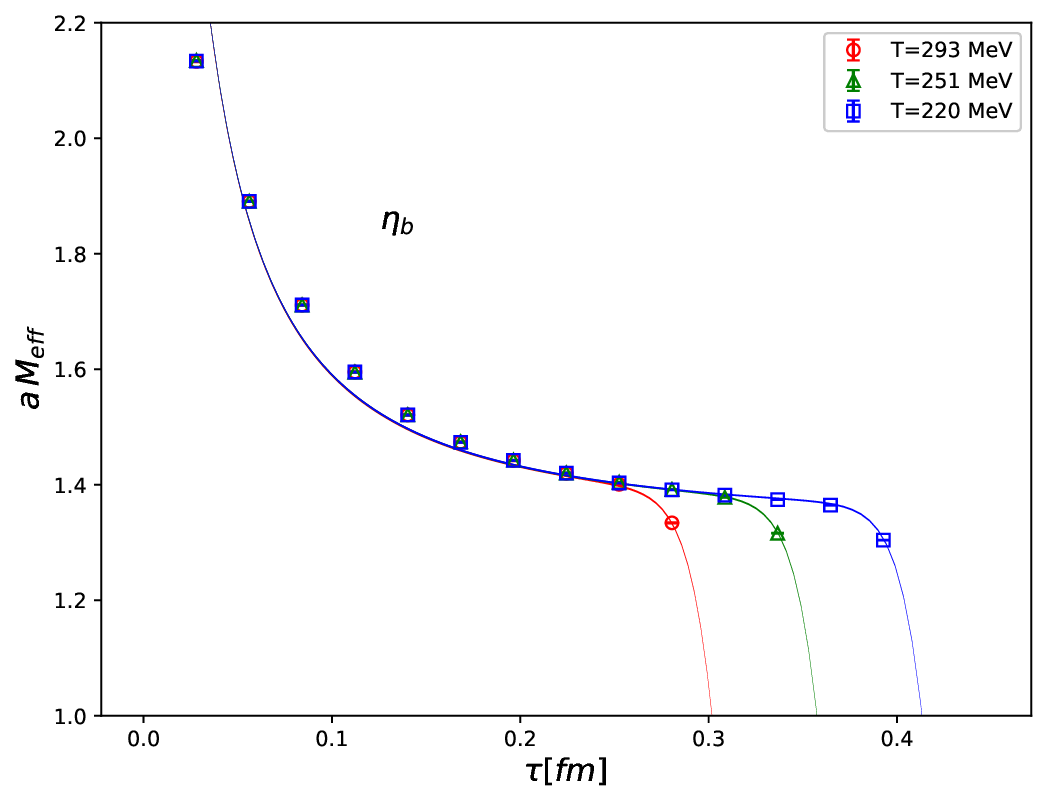}
	\caption{The effective mass, $M_{eff}$, is compared with lattice data and the corresponding spectral function in \Cref{sp_temp}: left panel for $\eta_c$ and right panel for $\eta_b$.	}
    \label{eff_mass}
\end{figure*}

\begin{figure*}
    \centering
    \includegraphics[width=8cm]{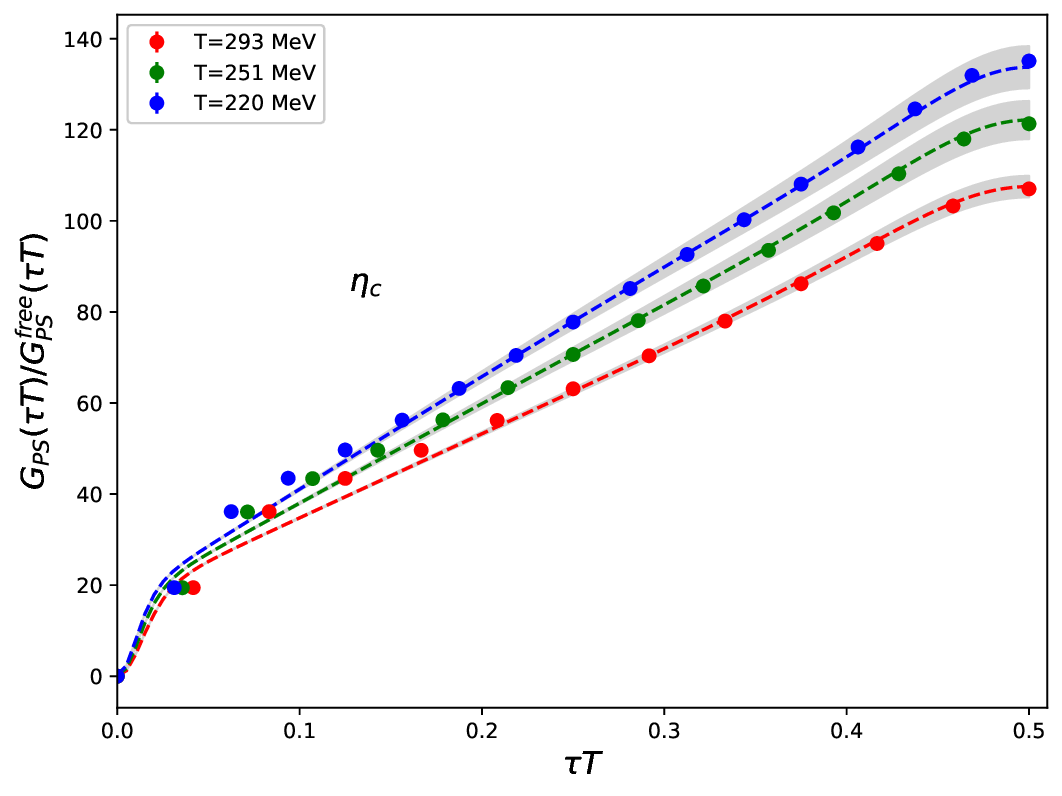}
    \includegraphics[width=8cm]{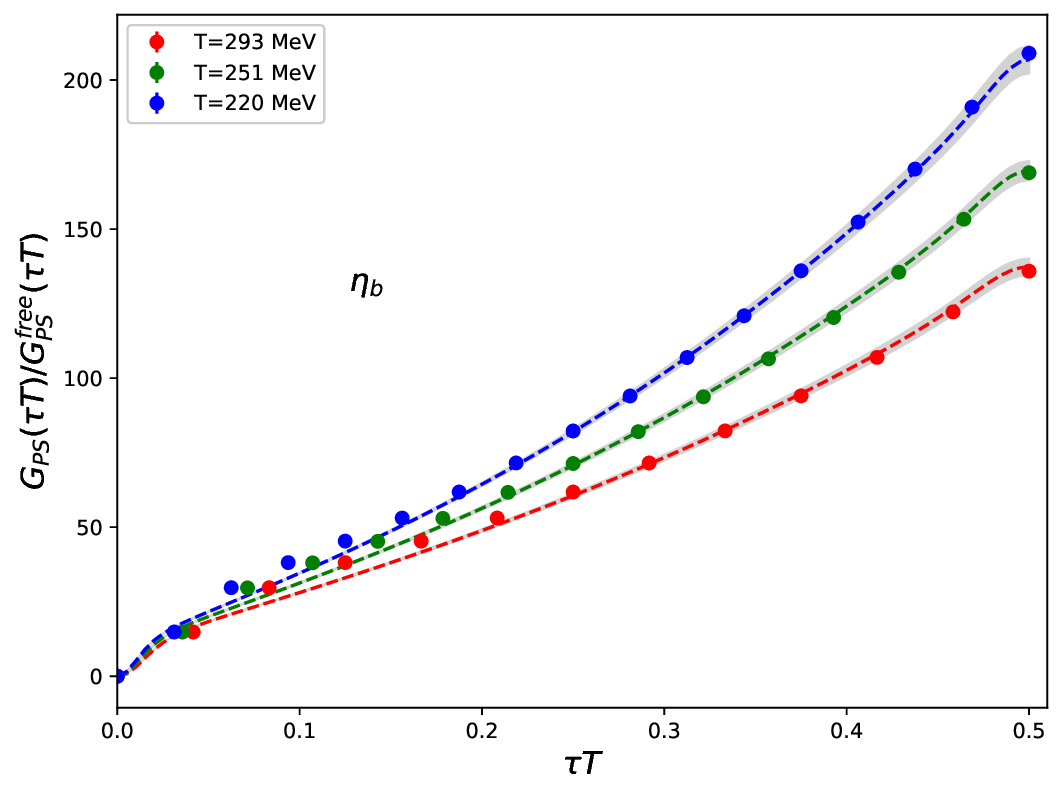}
	\caption{Prediction of the pseudoscalar correlator from the spectral function in \Cref{Spec fn}, compared with the lattice correlator using relativistic quarks: left panel for $\eta_c$, right panel for $\eta_b$.
	}
    \label{Param_lattice}
\end{figure*}

To get an estimate of the width, we fit the spectral function around the peak using the skewed Breit-Wigner form \cite{Burnier:2015tda}:

\begin{equation} 
\rho(\omega \approx M) = A \frac{\Gamma/2}{\Gamma^2/4 + (\omega - M)^2} + 2\delta \frac{(\omega - M) \Gamma}{(\omega - M)^2 + \Gamma^2} + \ldots
\label{sp_param}
\end{equation}

This form is motivated by the fact that thermal interactions shift the pole of the retarded propagator from the real axis into the complex plane, resulting in a spectral function (the imaginary part of the retarded correlator) with a Breit-Wigner structure near the peak.

The fit for the spectral function is illustrated in \Cref{Param_sp}. The fitted width, $\Gamma(1S)$, and the thermal mass, $M$, obtained from the peak position of the spectral function are shown in \Cref{width}. The error bars in the width and mass arise from the uncertainty in the estimates of the zero-temperature mass and the imaginary part, as discussed in \Cref{app_b}. As can be seen from the figure, the thermal decay width increases with temperature. The results suggest a width for the $\eta_c(1S)$ in the range $\Gamma \sim 300 - 600$ MeV, while the $\eta_b(1S)$ exhibits a significantly smaller decay width, approximately 10–15 times smaller, in the range $\Gamma \sim 20 - 60$ MeV. 

The thermal mass of the $\eta_c(1S)$ state is consistent with the zero-temperature mass within error bars, while the thermal mass of the $\eta_b(1S)$ state shows a departure from the zero-temperature mass, decreasing with increasing temperature. As mentioned earlier, this is due to the fact that the real part of the potential always tends to shift the peak position toward smaller values, whereas the imaginary part pushes the peak position toward larger values. This suggests that the impact of the imaginary part on the charm state is much larger than that of the real part in the temperature range we considered, and that $\eta_c(1S)$ is already close to its melting temperature at our highest temperature $293\, \text{MeV}$. In contrast, the impact of the real part on the bottom state is much larger compared to the imaginary part, resulting in a downward shift with increasing temperature.

\section{Comparison with lattice}
\label{check}
As the pseudoscalar spectral function does not contain any additional structure in the $\omega \sim 0$ region we can use the spectral function from \Cref{sp_temp} to describe the long-distance part of the lattice correlator in \Cref{def.pt}.  

On the lattice, we directly calculate \( G^{\s{\text{PS}}}(\tau T)/M_{\s{B}}^2 \), while the spectral function we calculated is \( \rho^{\s{\text{PS}}}(\omega)/m(\mu_{\s{\text{ref}}})^2 \). A multiplicative renormalization factor is needed for the lattice correlator to transform it to the $\overline{\text{MS}}$ scheme at the mass scale $\mu = 1/a$ , and subsequently multiply by the factor $\left(\frac{m(\mu=1/a)}{m(\mu_{\text{ref}})}\right)^2$. However, this renormalization factor is currently not known for 2+1 flavor QCD. Therefore, for comparison we calculate the following quantity from the lattice QCD data:  

\begin{equation}
a\,M_{\text{eff}}(\tau_{i}) = \log\left( \frac{G^{\s{E}}_{\s{\text{PS}}}(\tau_{i})}{G^{\s{E}}_{\s{\text{PS}}}(\tau_{i+1})} \right).
\end{equation}  

In this ratio, the renormalization constant cancels out, allowing for a direct comparison between the lattice data and the correlation function obtained from the spectral function in \Cref{sp_temp}. This comparison is shown in \Cref{eff_mass}, where we also include the error of the spectral function in the correlator. From this, we observe that the correlator obtained from the spectral function in \Cref{sp_temp} is consistent with lattice QCD data at large $\tau$ values within error bars across all temperatures.  

To quantify the level of agreement, we calculate the following $\chi^2$, defined as:

\begin{equation}
\chi^2 = \sum_{\tau > \tau_{min}} \frac{(M_{\s{\text{eff}}}^{\s{\text{data}}} - a M_{\s\text{eff}}^{\s{\text{prediction}}})^2}{(\text{err}_{\s\text{eff}}^{\s{\text{data}}} + \text{err}_{\s\text{eff}}^{\s{\text{prediction}}})^2},
\end{equation}

Where $\text{err}_\text{eff}^{\text{data}}$ represents the statistical error on the lattice effective masses, and $\text{err}_\text{eff}^{\text{prediction}}$ is the systematic uncertainty in the correlator from the spectral function. For the $\eta_c$ and $\eta_b$ channels, we find that $\tau_{min}=0.22\,\text{fm}$ and $\tau_{min}=0.28\,\text{fm}$ give a $\chi^2/\text{ndf}$ range of $0.1-0.9$. The value of $\tau_{min}$ is expected to be larger for the $\eta_b$ state than for the $\eta_c$ state as the bottom quark mass is much larger than the charm quark mass, leading to larger lattice artifacts at larger distances for the $\eta_b$ case. 

\Cref{eff_mass} also indicates that it should be possible to fit the lattice correlator at all temperatures using the spectral function with a single multiplicative constant. The overall conversion factor that was mentioned before can be absorbed into the overall temperature-independent fitting coefficient. Since the prediction also contains errors, we minimize the \(\chi^2\) function, defined as:

\begin{equation}
\chi^2(A) = \sum_{T} \sum_{\tau > \tau_{\text{min}}} \frac{(\text{corr}^{\s{\text{data}}}(T) - A\,\text{corr}^{\s{\text{prediction}}}(T))^2}{(\text{err}_{\s{\text{corr}}}^{\s{\text{data}}}(T) + A\,\text{err}_{\s{\text{corr}}}^{\s{\text{prediction}}}(T))^2}
\end{equation}

This yield value of \( A \sim 11 \) for the \( \eta_c \) correlator and \( A \sim 7 \) for the \( \eta_b \) correlator. 

In \Cref{Param_lattice}, we have plotted \( G_{\s{\text{PS}}}/G_{\s{\text{PS}}}^{\s{\text{free}}} \) from both, the direct lattice correlators and the correlators obtained from the spectral function. Here, we divide the correlator by \( G_{\s{\text{PS}}}^{\s{\text{free}}} \), the correlator in the free theory, to improve visualization. 

From \Cref{eff_mass} and \Cref{Param_lattice}, we observe that our spectral function, obtained from the Schr\"odinger equation using a screened thermal potential, provides a valid description of the \( \eta_c \) and \( \eta_b \) correlators obtained directly from the lattice. We therefore conclude that the non-perturbatively determined complex potential, with a real part that describes color screening of the QGP, is consistent with all the lattice measurements of pseudoscalar charmonium and bottomonium correlators we studied.

\section{Conclusion}
\label{check}
In this paper we study quarkonium spectral functions in the pseudoscalar channel using a non-perturbatively determined thermal lattice potential and compared this to a calculation of the full relativistic pseudoscaler charmonium and bottomonium point-point correlators in lattice QCD. For the quark propagator we used clover-improved Wilson fermions, and for the background gauge field, we used 2+1 flavor HISQ configurations with a pion mass of $320 \,\text{MeV}$. For the spectral reconstruction process of this correlator, we combined the spectral function in different regions using physics-based information.

As pointed out, different regions of the spectral function can be computed using different methods. The perturbative vacuum spectral function was used for the region $\omega \gg 2M_q$, whereas for $\omega \sim 2M_q$, a non-perturbatively determined complex thermal potential was used in the Schr\"odinger equation to extract the spectral function. For the pseudoscalar channel, the combination of the spectral functions from these two regions provides a sufficient description.

The thermal complex potential was calculated from non-perturbatively determined Wilson line correlators, which itself has an analytic continuation problem. To extract the potential, an ansatz for the parameterization of the Wilson line correlators was used, motivated by the general $\tau$ dependence of the HTL spectral function, as shown in \Cref{Param_WL}. This guarantees the existence of the limit in \Cref{pot_def}. As demonstrated in \Cref{Fit}, the parametrization effectively describes the lattice QCD correlator within a reasonable range in the intermediate region. The $\tau \sim 0$ and $\tau \sim \beta$ regions are not expected to be described by this correlator, as the thermal potential originates from the long-time behavior of Wilson line correlator.

At long distances, the potential becomes quite noisy, so gradient flow at different depths was used to enhance the signal at large separations, as seen in \Cref{Flow_depth}. However, it was observed that the flow distorts the short-distance behavior of the real part of the potential. A smooth matching of the potentials at different flow depths was performed to obtain a reliable estimate for the real part. The results indicate the presence of color screening in the QGP phase, as shown in \Cref{Vre_temp}. We found that the imaginary part of the potential does not show significant flow dependence within error bars, as shown in \Cref{IM_Flow_depth}. Nevertheless, a zero flow-time extrapolation using a linear fit was carried out to obtain the final imaginary part. The imaginary part increases with both, distance and temperature, as illustrated in \Cref{Vim_temp}. 
Note that our distances are too small to observe the saturation of the imaginary part expected to occur at larger distances. 
For solving the Schr\"odinger equation in \Cref{schdn}, the real and imaginary parts of the potential were parameterized as in \Cref{real_param} and \Cref{imag_param}. The short-distance part of the real potential, considered temperature-independent, was replaced with the renormalon-subtracted perturbative vacuum potential, as demonstrated in \Cref{Param_real}

The spectral function was calculated using the potentials in \Cref{Spec fn}.  The quark pole mass was determined using the zero-temperature potential. By solving the Schr\"odinger equation, the spectral function in the threshold region was obtained and subsequently matched with the perturbative spectral function in the ultraviolet regime, as shown in \Cref{matching}. The resulting pseudo scalar quarkonia spectral functions at various temperatures are shown in \Cref{sp_temp}, representing one of the main results of this paper. It was observed that charmonium states are significantly more sensitive to temperature compared to bottomonium states. For charm quarks, thermal effects are predominantly influenced by the imaginary part of the potential, while for bottom quarks, the real part plays a more critical role in the temperature range considered.

In \Cref{check}, it is shown that these spectral functions can predict the lattice correlator calculated using the relativistic quark propagator. This agreement between the predicted and measured lattice correlators based on the thermal screened potential supports the conclusion that the screening of the QCD plasma is consistent with all lattice QCD data measured.

There are various ways to extend this work. We have studied the pseudoscalar channel; this analysis can be extended to the vector channel, where, in addition to the bound state region, one needs to model the transport part of the spectral function. Here, we use the thermal potential in the static limit; at high temperatures, the mass correction to the thermal potential becomes important. Therefore, relativistic corrections to the static potential also need to be calculated. We would also like to extend the formalism of this paper to small non-zero baryonic density, which will be relevant for low-energy RHIC and FAIR experiments.

\section{Acknowledgments}
The generation of gauge configurations and the measurement of Wilson line correlators were carried out using SIMULATeQCD \cite{HotQCD:2023ghu}. These configurations have also been used in other studies by the HotQCD collaboration \cite{Altenkort:2023oms, Bazavov:2023dci, Ali:2024xae}. Some of the Wilson line correlator data used in this paper was taken from~\cite{Bazavov:2023dci}, and we thank the authors for providing it. We also thank Luis Altenkort and Hai-Tao Shu for their efforts in integrating the mesonic correlator into the QUDA~\cite{Clark:2009wm} code, which was used for the measurement of charm and bottom-point correlation functions. We would like to thank Saumen Datta for various discussions and reading the manuscript. The authors 
acknowledge support by the Deutsche Forschungsgemeinschaft (DFG, German Research Foundation) through the CRC-TR 211 `Strong-interaction matter under extreme conditions' – project number 315477589 – TRR 211.
For the computational work we used the Bielefeld GPU cluster and the LUMI-G supercomputer. We acknowledge the EuroHPC Joint Undertaking forawarding  this  project  access  to  the  EuroHPC  supercomputer  LUMI-G, hosted by CSC (Finland) and the LUMI consortium through a EuroHPC Extreme Scale Access call.
\appendix
\section{Leading order perturbative calculation}
\label{appa}
\begin{widetext}
In this section, we will calculate the gradient-flowed Wilson line correlator at leading order to understand the effect of flow on the thermal potential. The Wilson line correlator, in terms of Wilson lines, is defined in Coulomb gauge as
\begin{equation*}
 W(r_{\s{L}},r_{\s{R}},\tau;\tau_{\s{F}}) =  \frac{1}{N_c}\langle \text{Tr}[U_{\s{L}}(r_{\s{L}},\tau_{\s{F}})U_{\s{R}}^{\dagger}(r_{\s{R}},\tau_{\s{F}})] \rangle_{\s{T}},
\end{equation*}
where $\langle O \rangle_{\s{T}}$ denotes the thermal average.

Here, $U_{\s{L}}$ and $U_{\s{R}}$ is the Wilson line at finte flow time at position $r_{\s{L}}$ (postion of quark) and $r_{\s{R}}$ (position of antiquark) given by,
\begin{equation*}
    U_{\s{L/R}}(r_{\s{L/R}},\tau_{\s{F}})=\exp\left(i\,g \int_{0}^{\infty} A_{\s{0}}(r_{\s{L/R}}, \tau;\tau_{\s{F}})\right).
\end{equation*}
Expanding up to $O(g^2)$,
\begin{align}
 W(\vec r_{\s{L}},\vec r_{\s{R}},\tau,\tau_{\s{F}})  &=1-\frac{g^2}{4 N_c}\int_{0}^{\tau}\int_{0}^{\tau} \langle A^{a}_{\s{0}}(\vec r_{\s{L}},\tau_{\s{L1}};\tau_{\s{F}})A^{a}_{\s{0}}(\vec r_L,\tau_{L2},\tau_{\s{F}}) \rangle_{\s{T}} d\tau_{\s{L1}} d\tau_{\s{L2}}\notag\\
 &- \frac{g^2}{4 N_c}\int_{0}^{\tau}\int_{0}^{\tau} \langle A^{a}_{\s{0}}(\vec r_{\s{R}},\tau_{\s{R1}},\tau_{\s{F}})A^{a}_{\s{0}}(\vec r_{\s{R}},\tau_{\s{R2}},\tau_{\s{F}}) \rangle_{\s{T}} d\tau_{\s{R1}} d\tau_{\s{R2}} \notag \\
 &+\frac{g^2}{2 N_c}\int_{0}^{\tau} \int_{0}^{\tau} \langle A^a_{\s{0}}(\vec r_{\s{L}},\tau_{\s{L}},\tau_{\s{F}}) A^a_{\s{0}}(\vec r_{\s{R}},\tau_{\s{R}};\tau_{\s{F}}) \rangle_{T} d\tau_{\s{L}} d\tau_{{\s{R}}}.
 \label{wL_pert}
\end{align}

where $\tau_{\s{L}}$ and $\tau_{\s{R}}$ denote the time coordinate of the flowed gauge field at $r_{\s{L}}$ and $r_{\s{R}}$, respectively.

We denote the position space propagator as,
\begin{equation*}
D^{ab}(r_1-r_2,\tau_1-\tau_2;\tau_{\s{F}})= \langle A_{\s{0}}^{a}(r_1,\tau_1;\tau_{\s{F}})A_{\s{0}}^{b}(r_2,\tau_2;\tau_{\s{F}}) \rangle.
\end{equation*}
At finite flow time $\tau_{\s{F}}$, the propagator in the leading order is given by
\begin{equation*}
D_{\s{00}}^{ab}(\vec r, \tau;\tau_{\s{F}})=T\int \frac{d^3 \vec p}{(2\pi)^3} \sum_{n} \exp(i\omega_{n}\tau-i \vec p.\vec r) \exp[-2\tau_{\s{F}}(\omega_n^2+p^2)] D_{\s{00}}^{ab}(\omega_{n},\vec p),
\end{equation*}
where $\omega_{n}=2\pi n T$ is the Matsubara frequency. $D^{ab}(\omega_n,\vec p)$ is the standard HTL momentum space propagator in the Coulomb gauge given by,
\begin{equation*}
D_{\s{00}}^{ab}(\omega_n,p)=\frac{\delta^{ab}}{\omega_n^2+p^2+\Pi_{\s{E}}(\omega_n,p)}\frac{\omega_n^2+p^2}{p^2}.
\end{equation*}
Substituting this into \Cref{wL_pert} and performing the integration over $\tau_{L/R}$, we obtain,
\begin{align*}
W(r=|\vec r_{\s{L}}-\vec r_{\s{R}}|,\tau;\tau_{\s{F}}) &= 1-\frac{g^2}{2 N_c}\sum_{n}\int_{p}  \exp[-2\tau_{\s{F}}(\omega_n^2+p^2)]\frac{2-\exp(i\omega_n \tau)-\exp(-i\omega_n\tau)}{\omega_n^2} D_{\s{00}}^{aa}(\omega_n,\vec p) \\
 &+\frac{g^2}{2 N_c}\sum_{n}\int_{p}  \exp[-2\tau_{\s{F}}(\omega_n^2+p^2)] \exp(-i\vec p.\vec r)\frac{2-\exp(i\omega_n \tau)-\exp(-i\omega_n\tau)}{\omega_n^2} D_{\s{00}}^{aa}(\omega_n,\vec p).
\end{align*}
Simplifying, we get
\begin{align}
W(r,\tau,\tau_{\s{F}}) &= 1- g^2 \int \frac{d^3 \vec p}{(2\pi)^3}\,(1-\exp(i\vec p.\vec r))\, \exp(-2\tau_{\s{F}} p^2)\,I(\tau,\tau_{\s{F}},p),
\end{align}
where
\begin{align}
I(\tau,\tau_{\s{F}}, p) &= \frac{1}{2 N_c}\sum_{n} \exp(-2\tau_{\s{F}}\omega_n^2)\frac{2-\exp(i\omega_n \tau)-\exp(-i\omega_n\tau)}{\omega_n^2} D^{aa}(\omega_n,\vec p).
\end{align}

Separating out the $\omega_n=0$ mode and using the propagator’s spectral representation
\begin{equation*}
\frac{1}{\omega_n^2+p^2+\Pi_{\s{E}}(\omega_n,p)} =  \int_{-\infty}^{\infty } \frac{dp_0}{\pi} \frac{\rho_{\s{E}}(p_{\s{0}},\vec p)}{p_{\s{0}}-i\omega_n},
\end{equation*}
$\Pi_{\s{E}}(\omega_n,\vec p)$ is the longitudinal gluon self energy.
we obtain
\begin{align*}
I(\tau,\tau_{\s{F}},p) &= g^2 C_{\s{F}}\Bigg[ \frac{\tau ^2 T}{p^2 +\Pi_{\s{E}}(0,\vec p)}+\int \frac{dp_{\s{0}}}{\pi} F(\tau,\tau_{\s{F}},p_{\s{0}},p) \rho_{\s{E}}(p_{\s{0}},\vec p) \Bigg],
\end{align*}
where $F(\tau,\tau_{\s{F}},p_{\s{0}},p)$ contains the following Matsubara sum,
\begin{align*}
F(\tau,\tau_{\s{F}},p_{\s{0}},p)=p_{\s{0}} T \sum_{n\neq0}  (2-\exp(i\omega_n \tau)-\exp(-i\omega_n\tau))\exp(-2 \omega_n^2 \tau_{\s{F}})\left(\frac{1}{\omega_n^2}+\frac{1}{\vec p^2}\right) \frac{1}{p_{\s{0}}^2+\omega_n^2}.
\end{align*}
To perform the Matsubara sum, we apply the Laplace transformation to the above equation with respect to \(\tau_{\s{F}}\):

\[
F(s,p_{\s{0}},p) = \mathcal{L}[F(\tau_{\s{F}},p_{\s{0}},p)] = \int_{0}^{\infty} F(\tau_{\s{F}},p_{\s{0}},p) \, e^{-s \tau_{\s{F}}} \, d\tau_{\s{F}}.
\]

This leads to:

\[
F(s,p_{\s{0}},p) = p_{\s{0}} T \sum_{n\neq0}  \left(2 - e^{i\omega_n \tau} - e^{-i\omega_n\tau} \right) \frac{1}{s+2\omega_n^2} \left(\frac{1}{\omega_n^2} + \frac{1}{\vec{p}^2} \right) \frac{1}{p_{\s{0}}^2+\omega_n^2}.
\]

In the Laplace domain, the Matsubara sum can be easily evaluated using partial fractions. Performing the Matsubara sum, followed by the inverse Laplace transform and some algebraic simplifications, leads to the following expression:
\begin{align}
I(\tau,\tau_{\s{F}},p)&= C_{\s{F}} \Bigg[ \frac{\tau  }{p^2 +\Pi_{\s{E}}(0,\vec p)}+\int \frac{dp_{\s{0}} }{\pi} \mathcal{L}^{-1}\Bigg \{ p_{\s{0}} \frac{R(\tau,\sqrt{s/2})}{(s-2 p_{\s{0}}^2)}\biggl(\frac{2}{s}-\frac{1}{p^2}\biggr)+\frac{p_{\s{0}}\,R(\tau,p_{\s{0}})}{(s-2 p_{\s{0}}^2)} \biggl(\frac{1}{p^2}-\frac{1}{p_{\s{0}}^2}\biggr)\Bigg \}  \rho_{\s{E}}(p_{\s{0}},\vec p) \Bigg].
\end{align}
Here, $R(\tau,p_{\s{0}})$ is a periodic function given by:
\begin{align*}
R(\tau, p_{\s{0}}) = \frac{1+\exp(\beta p_{\s{0}})-\exp((\beta-\tau)p_{\s{0}})-\exp(\tau p_{\s{0}})}{p_{\s{0}}(\exp(\beta p_{\s{0}})-1)}.
\end{align*}
Using the convolution theorem, we can write:
\begin{align}
I(\tau, \tau_{\s{F}},p)&= C_{\s{F}} \Bigg[ \frac{1  }{p^2 +\Pi_{E}(0,\vec p)}+
  \int \frac{dp_{\s{0}} }{\pi} \Bigg(\mathcal{L}^{-1}\biggl\{\frac{p_{\s{0}}}{(s-2 p_{\s{0}}^2)}\biggl(\frac{2}{s}-\frac{1}{p^2}\biggr)\biggr\} *\mathcal{L}^{-1}\{R(\sqrt{s/2},\beta,\tau)\} \notag \\
  &+p_{\s{0}} \exp(2 p_{\s{0}}^2 \tau_{\s{F}}) R(p_{\s{0}},\beta,\tau)\biggl(\frac{1}{p^2}-\frac{1}{p_{\s{0}}^2}\biggr)\Bigg)  \rho_{\s{E}}(p_{\s{0}},\vec p)\Bigg].
\end{align}

We further refine this to obtain:
\begin{align}
I(\tau, \tau_{\s{F}},p) &=C_{\s{F}} \Bigg[ \frac{\tau}{p^2 +\Pi_{\s{E}}(0,\vec p)}\nonumber \\
  & + \int \frac{dp_{\s{0}} p_{\s{0}}}{\pi} \bigg \{ g(\tau_{\s{F}})*Q(\tau_{\s{F}},\beta,\tau)+ \exp(2 p_{\s{0}}^2 \tau_{\s{F}}) R(p_{\s{0}},\beta,\tau)\biggl(\frac{1}{p^2}-\frac{1}{p_{\s{0}}^2}\biggr)\bigg \}  \rho_{\s{E}}(p_{\s{0}},\vec p)\Bigg].
\label{I_fl}
\end{align}

Here,
\begin{align*}
 g(\tau_{\s{F}})=\mathcal{L}^{-1}\biggl\{\frac{1}{(s-2 p_{\s{0}}^2)}\biggl( \frac{2}{s}-\frac{1}{p^2}\biggr)\biggr\}&=-\frac{1}{p_{\s{0}}^2}-\biggl(\frac{1}{p^2}-\frac{1}{p_{\s{0}}^2}\biggr) \exp(2 p_{\s{0}}^2 \tau_{\s{F}}),   \\
 Q(\tau_{\s{F}},\beta,\tau)&=\mathcal{L}^{-1}\{R(\sqrt{s/2},\beta,\tau)\}.
\end{align*}

The following convolution can be further simplified:
\begin{align*}
g(\tau_{\s{F}})*Q(\tau_{\s{F}},\beta,\tau)&=\int_{0}^{\tau_{\s{F}}}\, d\nu\, Q(\nu,\beta,\tau)\,g(\tau_{\s{F}}-\nu)\\
 &= - \bigg[ \frac{1}{p_{\s{0}}^2} \ P(0,\tau_{\s{F}},\beta,\tau) + \biggl(\frac{1}{p^2}-\frac{1}{p_{\s{0}}^2}\biggr) \exp(2 p_{\s{0}}^2 \tau_{\s{F}})\, P(2 p_{\s{0}}^2,\tau_{\s{F}},\beta,\tau)\bigg].
\end{align*}
Here, $P(2 p_{\s{0}}^2,\tau_{\s{F}},\beta,\tau)=\int_{0}^{\tau_{\s{F}}}\, d\nu\, Q(\nu,\beta,\tau)\exp(-2 p_{\s{0}}^2 \nu)$. 

If $\tau_{\s{F}} \to \infty$, then $P(2 p_{\s{0}}^2,\tau_{\s{F}},\beta,\tau)$ corresponds to the Laplace transform of $Q$ at the point $2 p_{\s{0}}^2$. Since $Q$ is already the inverse Laplace transform of $R(\sqrt{s/2},\beta,\tau)$, performing the Laplace transform will lead us back to $R(\sqrt{s/2},\beta,\tau)$ at the point $2 p_{\s{0}}^2$. 
To generalize this, we write:
\begin{align*}
P(2 p_{\s{0}}^2,\tau_{\s{F}},\beta,\tau)&=\int_{0}^{\infty}\,\theta(\tau_{\s{F}}-\nu)\, Q(\nu,\beta,\tau)\exp(-2 p_{\s{0}}^2 \nu)\,d\nu\\
                            &=\int_{0}^{\infty}d\nu \int_{-\infty}^{\infty} dz\frac{1}{2\pi i} \frac{1}{z-i\epsilon} \exp(i(\tau_{\s{F}}-\nu) z) Q(\nu,\beta,\tau)\exp(-2 p_{\s{0}}^2 \nu)\\
                            &= \int_{-\infty}^{\infty} dz\frac{\exp(i \tau_{\s{F}} z)}{2\pi i(z-i\epsilon)} \int_{0}^{\infty}d\nu \exp[-(2 p_{\s{0}}^2+i z)\nu]Q(\nu,\beta,\tau)\\
                            &= \int_{-\infty}^{\infty} dz\frac{\exp(i \tau_{\s{F}} z)}{2\pi i(z-i\epsilon)} R\biggl(\sqrt{p_{\s{0}}^2+i \frac{z}{2}},\beta,\tau\biggr).
\end{align*}

Substituting everything back into $I(\tau,p_{\s{0}},p)$ in \Cref{I_fl}, we obtain:
\begin{align*}
I(\tau,\tau_{\s{F}},p) &= C_{\s{F}} \Bigg[ \frac{\tau-P(0,\tau_{\s{F}},\beta,\tau)}{p^2 +\Pi_{\s{E}}(0,\vec p)}\\
  & + \int \frac{dp_{\s{0}} p_{\s{0}}}{\pi} \biggl(\frac{1}{p^2}-\frac{1}{p_{\s{0}}^2}\biggr)\exp(2 p_{\s{0}}^2 \tau_{\s{F}}) \bigg \{ R(p_{\s{0}},\beta,\tau)-P(2 p_{\s{0}}^2,\tau_{\s{F}},\beta,\tau)\bigg \}  \rho_{\s{E}}(p_{\s{0}},\vec p)\Bigg].
\end{align*}

The Wilson line correlator is then,
\begin{align}
W(r,\tau;\tau_{\s{F}}) &= 1- g^2\int \frac{d^3 \vec p}{(2\pi)^3}(1-\exp(i\vec p.\vec r)) \exp(-2 \tau_{\s{F}} p^2)I(\tau, \tau_{\s{F}},p)+...\nonumber\\
&\approx \exp\left(-g^2\int \frac{d^3 \vec p}{(2\pi)^3}(1-\exp(i\vec p.\vec r)) \exp(-2 \tau_{\s{F}} p^2)I(\tau, \tau_{\s{F}},p)\right)  
\label{W_final_p}
\end{align}
From this equation, we can see that our ansatz for fitting the Wilson line correlator in \Cref{Param_WL} also holds in the case of leading-order flowed configurations. Specifically, we observed that $\log(W(r,\tau))$ can be decomposed into a combination of a linear $\tau$ term and a periodic $\tau$ term. At zero flow time \Cref{W_final_p} reproduces the expression given in \cite{Laine:2006ns}.  

To see the effect of the flow on this correlator, we compute the above expression numerically and calculate the effective mass, defined as  
\begin{equation}
    m_{eff}(r,\tau_i;\tau_{\s{F}}) = \log\left(\frac{W(r,\tau_{i},\tau_{\s{F}})}{W(r,\tau_i+a,\tau_{\s{F}})}\right),
\end{equation}  
where $a$ is the lattice spacing.  

We use the two-loop thermal coupling $g^2(T)$ and the Debye mass $m_{\s{D}}^2 = \Pi_{\s{E}}(0,p)$ from \cite{Laine:2005ai}. The renormalization scale is chosen as $\mu_{min} = 9.082 T$, corresponding to the minimum sensitivity scale. The spectral function $\rho_{\s{E}}(p_{\s{0}},p)$ is taken from \cite{Laine:2006ns} and is given by  
\begin{align}
    \rho_{\s{E}}(p_{\s{0}},p) &= \pi \, \text{sign}(p_{\s{0}}) \delta(p_{\s{0}}^2 - p^2 - \text{Re}[\Pi_{\s{E}}(p_{\s{0}},p)]), \quad p < p_{\s{0}}, \\
    \rho_{\s{E}}(p_{\s{0}},p) &= -\frac{\pi m_{\s{D}}^2 p_{\s{0}}}{2 |\vec{p}| (p^2 + m_{\s{D}}^2)^2}, \quad p > p_{\s{0}}.
\end{align}  
The momentum cutoff in the integrations in \Cref{W_final_p} is set to $\pi/a$.  
\begin{figure*}
    \centering
    \includegraphics[width=8cm]{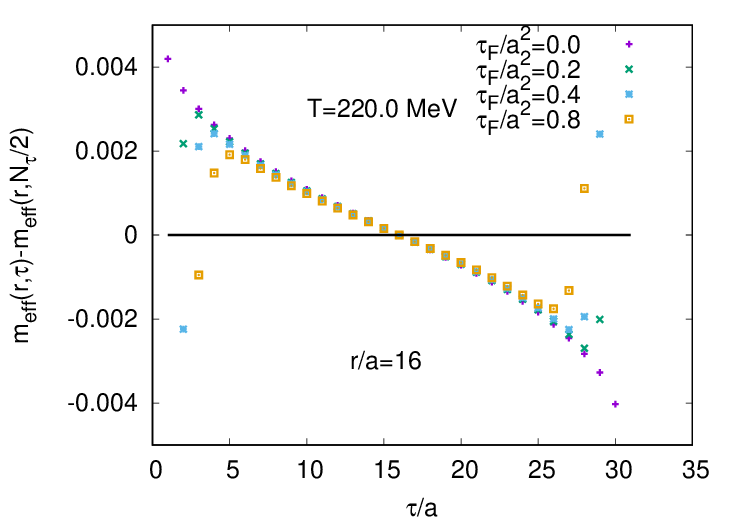}
    \includegraphics[width=8cm]{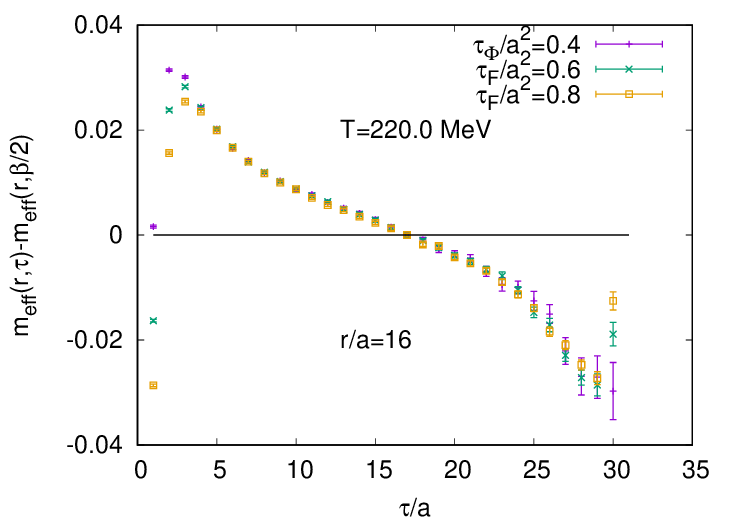}
	\caption{The flow time dependence of the effective mass at $T=220\,\text{MeV}$ and $r/a=16$. (Left) LO HTL perturbative result  (Right) Lattice QCD data. }
    \label{FT_LO}
\end{figure*}

In \Cref{FT_LO}, we plot $m_{eff}(r,\tau) - m_{eff}(r,N_{\tau}/2)$ as a function of $\tau$ for various flow times at $T = 220\,\text{MeV}$ ($N_\tau = 32$). The subtraction with respect to mid-point $N_{\tau}/2$ removes the flow-time-dependent additive renormalization constant. We observe that the flow time has a significant effect in the regions $\tau \sim 0$ and $\tau \sim \beta$. As the flow time increases, larger regions are affected. However, in the intermediate region, the effect of the flow remains small. On the right panel, we show $m_{eff}(r,\tau)$ from lattice data, which qualitatively exhibits similar behavior.  

\begin{figure*}
    \centering
    \includegraphics[width=8cm]{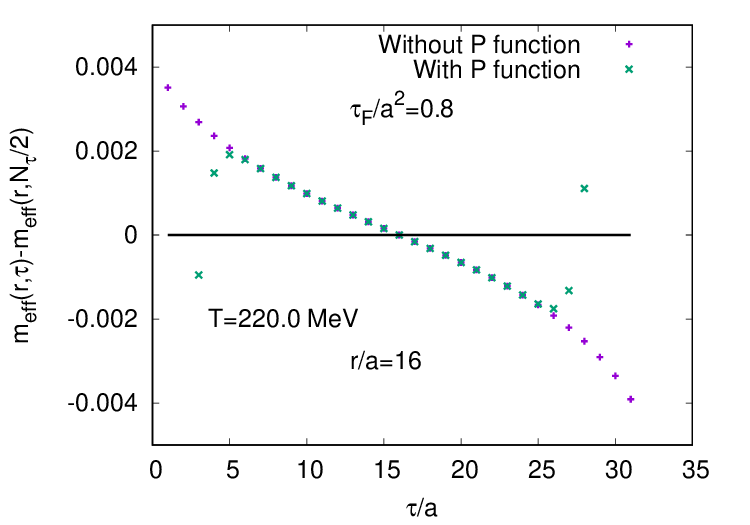}
	\caption{Comparison of LO effective mass with and without $P$ function}
	\label{wpo}
\end{figure*}

To compute the potential, we first need to perform analytic continuation $\tau \rightarrow i t$ and then evaluate  
$\lim_{t \rightarrow\infty} i \frac{\partial \log(W(r,it;\tau_{\s{F}}))}{\partial t}.$
The analytic continuation of the $P$ term in \Cref{W_final_p} turns out to be complicated. However the effects of this should be small in the determiniation of the potential. This can alrady be seen from the plot in \Cref{wpo} where we compare the effective mass with and without the $P(p_{\s{0}},\tau_{\s{F}},\beta,\tau)$ term in \Cref{W_final_p} for various flow times. This shows that this term primarily affects the regions $\tau \sim 0$ and $\tau \sim \beta$, while its influence in the intermediate region is negligible. Furthermore, the $P(p_{\s{0}},\tau_{\s{F}},\beta,\tau)$ term is entirely responsible for the non-monotonic behavior of the effective mass near $\tau \sim 0$ and $\tau \sim \beta$. Since the effect of $P(p_{\s{0}},\tau_{\s{F}},\beta,\tau)$ is small in the intermediate region, and the potential is extracted from this region, we neglect its contribution in the potential computation. The potential can then be easily obtained from the remaining terms, as given by:
\begin{figure*}
    \centering
    \includegraphics[width=8cm]{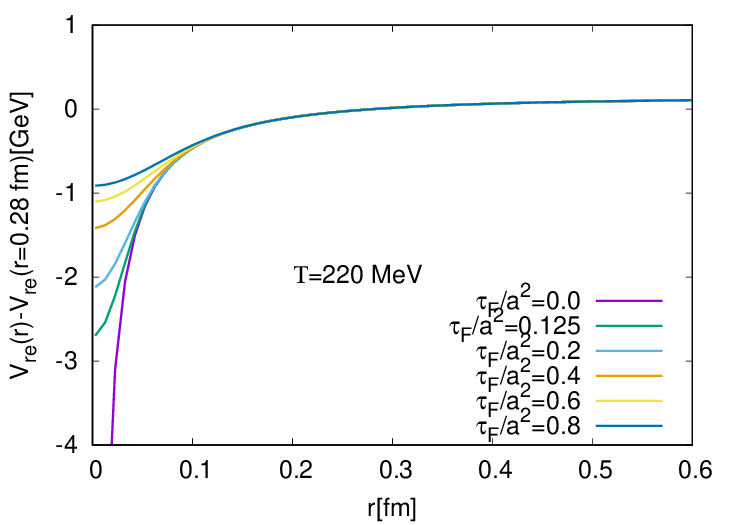}
    \includegraphics[width=8cm]{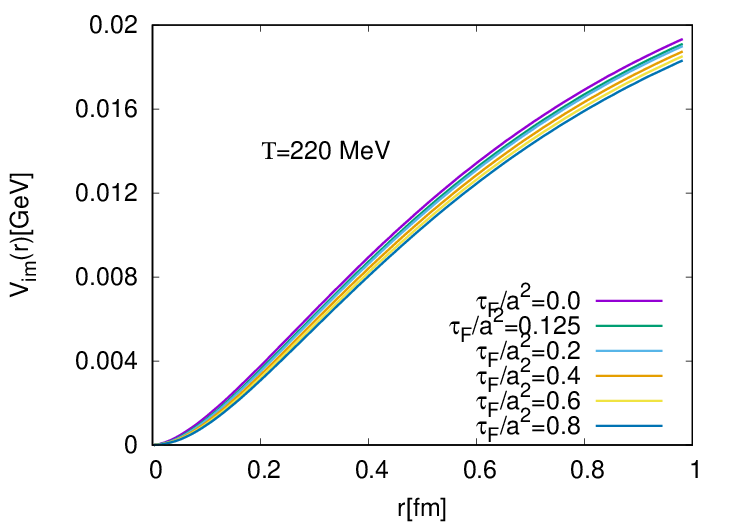}
	\caption{The flow time dependence of the thermal potential at LO. (Left) Real part (Right) Imaginary	part.}
    \label{FT_LO}
\end{figure*}

\begin{equation}
V_{\s{T}}^{\text{re}}[\tau_{\s{F}}, m, r] = C_{\s{F}} g^2 \int_{-\infty}^{\infty} \frac{d^3 \vec p}{(2\pi)^3} (\exp(i\vec p.\vec r)-1)\frac{\exp(-2\tau_{\s{F}} p^2)}{p^2+m_{\s{D}}^2} 
\end{equation}
\begin{equation}
V_{\s{T}}^{\text{im}}[\tau_{\s{F}}, m, r] = T \, C_{\s{F}}  m_{\s{D}}^2 g^2 \int_{-\infty}^{\infty} \frac{d^3\vec p}{(2\pi)^3}(\exp(i\vec p.\vec r)-1)\frac{\exp(-2\tau_{\s{F}} p^2)}{|\vec p|(p^2+m_{\s{D}}^2)^2}.
\end{equation}
This potential is plotted in \Cref{FT_LO} at $T=220 \,\text{MeV}$ for various flow times. In the left panel, we observe that the flow time dependence of the LO potential at short distances is quite similar to that of the non-perturbative potential in \Cref{Flow_depth}. Here, too, the effect of flow is significant at short distances compared to larger distances. However, the impact of flow on the LO real part is much stronger than what we observed non-perturbatively in \Cref{Flow_depth}.

In the right panel, we see that the imaginary part exhibits mild flow time dependence at all distances, with flow reducing its magnitude. However, this weak dependence is within the error bars of our lattice data and, therefore, cannot be resolved as shown in \Cref{IM_Flow_depth}.
\end{widetext}

\section{Systematic Uncertainly in the spectral function}
\label{app_b}
\begin{figure*}
    \centering
    \includegraphics[width=8cm]{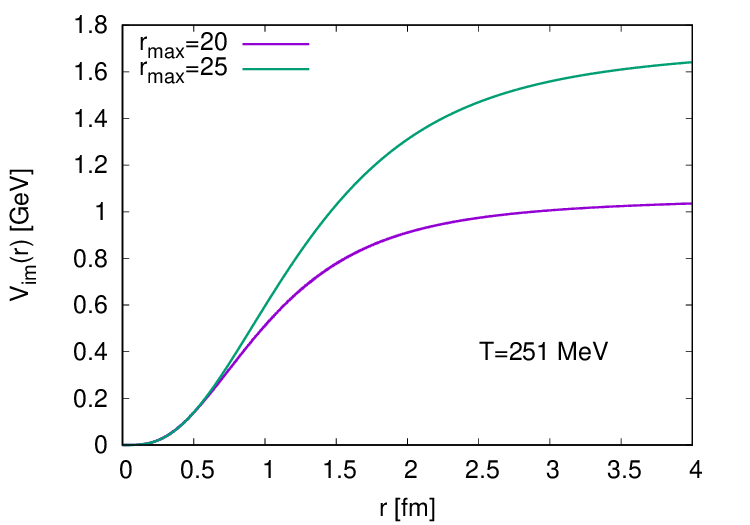}
    \includegraphics[width=8cm]{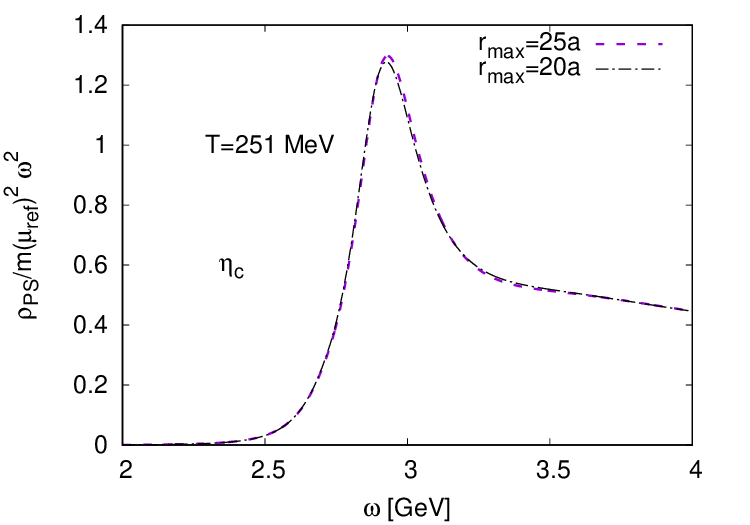}  
    \includegraphics[width=8cm]{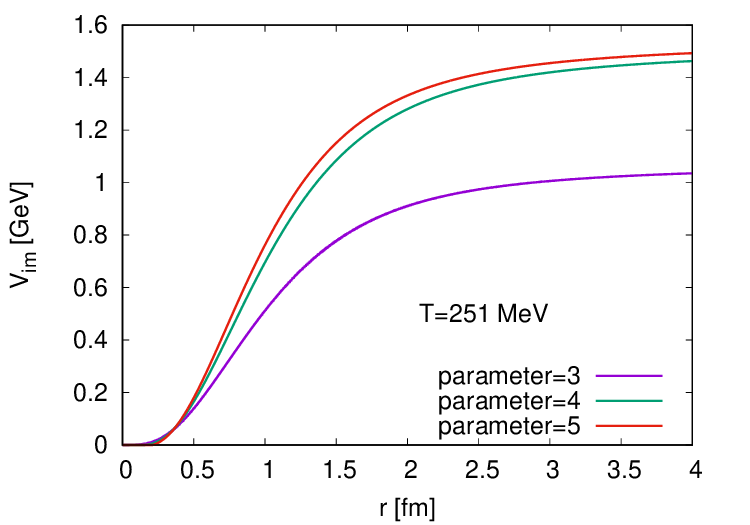}
    \includegraphics[width=8cm]{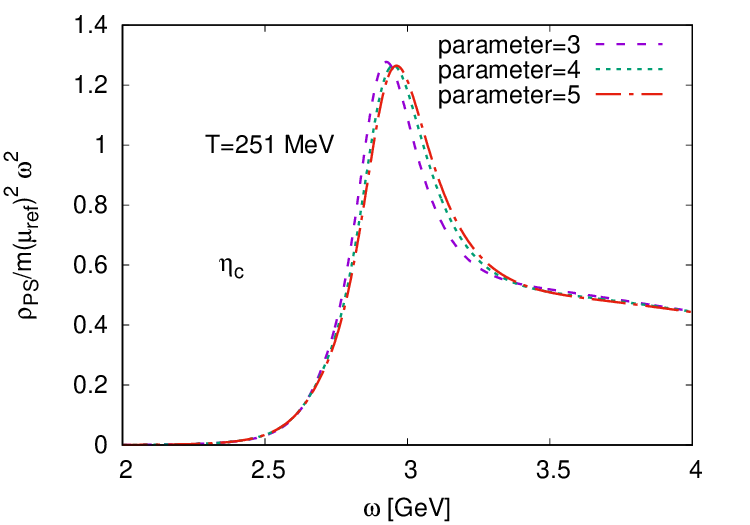}
	\caption{
Dependence of the spectral function on the imaginary part at \( T = 251 \) MeV for \( \eta_c \). Top panel: The right panel shows the dependence of the spectral function on the imaginary part, fitted using the parametrization in \Cref{im_param} from the range \( r_{\text{max}} = 20\,a \) to \( r_{\text{max}} = 25\,a \), as depicted in the left panel.Bottom panel: The right panel illustrates the dependence of the spectral function on the imaginary part obtained using different fit parameters from \Cref{Param_WL}, as shown in the left panel.	}
\label{im_sp}
\end{figure*}

In this section, we discuss the systematic uncertainty in the estimate of the spectral function in \Cref{sp_temp}, as well as the width and mass shift in \Cref{width}. As mentioned earlier, most of the error arises from the tuning of zero-temperature lattice masses, as shown in \Cref{Lattice}, and the imaginary part. Any uncertainty in the determination of the zero-temperature mass will introduce uncertainty in the determination of the pole quark mass and the additive renormalization of the potential, which, in turn, will affect the final spectral function and its characterization parameters. 

However, performing bootstrap error propagation for the spectral function is difficult in this case. As a result, we adopt a simpler approach. As mentioned earlier, we have fixed the bottom quark mass to \(4.78\, \text{GeV}\) and tuned the additive constant of the zero-temperature Cornell potential such that the ground state mass matches the lattice-determined \(\eta_{b}(1S)\) mass. To account for the error, we performed the matching for three values: \(M+\sigma_{\s{M}}\), \(M\), and \(M-\sigma_{\s{M}}\) for the \(\eta_{b}(1S)\) mass. This results in three different additive constants for the potential. The error in the \(\eta_{b}\) spectral function due to the uncertainty is thus arises through the uncertainty in these additive constants at each temperature. 

Next, we tune the charm quark mass using this additive constant by solving the Schr\"odinger equation for three values: \(M-\sigma_{\s{M}}\), \(M\), and \(M+\sigma_{\s{M}}\). Since there are three additive renormalizations, this results in nine charm quark mass values, contributing to the error in the \(\eta_c\) spectral function from zero temperature mass tuning. The average charm quark mass is found to be \(1.35 \pm 0.01\, \text{GeV}\). 
Now, we have to include the error of the thermal potential in the spectral function. We ignore the error on the real part because it is small compared to error introduced from other sources. On the other hand, the error in the imaginary part is significant, and therefore, we consider the effect of this uncertainty on the imaginary part.

The main source of uncertainty in the imaginary part comes from the number of fitting parameters in the ansatz given in \Cref{Param_WL}. We have also studied the systematic uncertainty by changing the fitting range and parameterization for each value of the number of parameters. The very short-distance region, around \(\sim 0.2\,\text{fm}\), cannot be fitted with this form, as in this region, \(V_{im}\) becomes very small and is consistent with zero. This very small value causes numerical accuracy problems in solving the Schr\"odinger equation. To avoid this, we have regularized the imaginary part as \(\max(10^{-4} M {\alpha}^2, V_{im}(r)\)). We have checked that the details of the parameterization at short distances do not have a significant impact on the spectral function.

In the top left panel of \Cref{im_sp} we have shown the imaginary part from 3 parameter fit for the fitting with $r_{min}=3\,a$ and $r_{max}=20\,a-25\,a$. On the right panel of \Cref{im_sp} we see although that the large distance part of this is very different, but nevertheless the effect of this difference on the spectral function is negligible.

In the bottom left panel of \Cref{im_sp}, we showd the imaginary part with different fit parameter range for the fit range with $r_{max}=20\,a$, and the corresponding spectral function on bottom right panel is shown to have a mild dependence on the the fit parameters. 

These two sources (mass tuning and imaginary part) of uncertainty we have taken into account when calculating the decay width and thermal mass from the spectral function \Cref{width} and also in the computation of Euclidean correlator for the comparsion with directly measured lattice correlaror in \Cref{eff_mass} and \Cref{Param_lattice}.

\newpage
\bibliography{ref}

\end{document}